\documentclass{article}

\usepackage{PRIMEarxiv}

\usepackage[utf8]{inputenc} 
\usepackage[T1]{fontenc}    
\usepackage{hyperref}       
\usepackage{url}            
\usepackage{booktabs}       
\usepackage{amsfonts}       
\usepackage{nicefrac}       
\usepackage{microtype}      
\usepackage{lipsum}
\usepackage{graphicx}
\graphicspath{{media/}}     

\title{Scanning Volcanoes by Synthetic Aperture Radar 
\thanks{\textit{\underline{Citation}}: 
\textbf{Authors. Title. Pages.... DOI:000000/11111.}} 
}

\author{
  Filippo Biondi \\
  University of Strathclyde \\
  Engineering Department \\
  Glasgow (UK)\\
  \texttt{\{Filippo Biondi\}filippo.biondi@strath.ac.uk} \\
}

\begin{document}
\maketitle

\begin{abstract}
A problem with synthetic aperture radar (SAR) is that due to the poor penetrating action of electromagnetic waves within solid bodies, the ability to observe through distributed targets is precluded. In this context, indeed, imaging is only possible on targets distribute on the scene surface.
\end{abstract}


\section{Introduction}
Vesuvius is an active volcano and dominates the central part of the Campanian coast, located in Italy. The morphology of the volcanic complex (including the Campi Flegrei) clearly reveals its active nature, characterised by eruptive cones and lava flows \cite{moe2002view}.  From a morphological point of view, Vesuvius is a volcanic complex composed of Mount Somma, whose activity ended with the formation of a summit caldera. The eruptive activity of this complex started more than 25000 years ago and gave rise to at least 5 Plinian eruptions, highly explosive, the most famous of which was the eruption of Pompeii (79 AD). This eruption buried the Roman cities of Pompeii, Stabia and Herculaneum and was described in two letters by Pliny the Younger to the Roman historian Tacitus, which are documents of fundamental importance for volcanology. All this has led to Vesuvius being characterised as one of the most dangerous volcanoes in the World \cite{sigurdsson1982eruption}. The reinterpretation of the volcanological and historical evidence shows that the eruption of Vesuvius in 79 AD consisted of two main phases. The initial Plinian phase, lasting 18-20 hours, caused a large pumice fallout, resulting in the slow accumulation of a layer of pumice up to 2.8 m thick over Pompeii and other regions to the south \cite{papale1993modeling}. Research \cite{santacroce1983general} gave the general model for the behaviour of the Soma-Vesuvius distinguishing three-classes in the eruptive pattern. The authors of \cite{auger2001seismic} gave the seismic evidence of an extended magmatic sill under Vesuvius. Seismic data were used to characterize its magmatic system. Authors found evidence of an extended (at least 400 km$^2$) low-velocity layer at about 8 km depth. The inferred seismic velocities (about 2.0 km/s), as well as other evidence, indicate an extended sill with magma interspersed in a solid matrix.
	
The presentation of the velocity structure of the Somma-Vesuvius volcano, obtained by joint inversion of P-and S-wave arrival times from both local earthquakes and shot data collected during the TOMOVES 1994 and 1996 experiments is presented in \cite{de2004seismicity}. The entire seismic catalog available from 1987 to 2000 (1003 events) was relocated in the 3-D velocity model obtained. Experimental imaging results show important implications for the hazard assessment of Somma-Vesuvius and all other volcanoes in the world.
The thermal model of the Vesuvius magma chamber, locating it at a depth of approximately 6 km has been studied in \cite{de2006thermal}. Given the high risk of the Vesuvius crater, it is essential to develop an effective forecasting model to measure the impact on the population of a possible Vesuvius eruption.
Efficient numerical modeling and the methodology adopted to evaluate the resistance of buildings under the combined action of volcanic phenomena have been described in \cite{zuccaro2008impact}.
Numerous works have been carried out in the field of sonic-inspired imaging, especially in the medical field, through the use of ultrasound. Theoretical studies made in the early 1980s suggest that ultrasonic imaging using the correlation technique can overcome some of the drawbacks of classical pulse echography.
An efficient high-resolution technique has been developed in \cite{benkhelifa1994echography} by transmitting coded signals using high-frequency transducers, up to 35–50 MHz, for ophthalmic echography to image fine eye structures.
The first high-frequency echographic images obtained with the prototype probe are presented in \cite{carotenuto2005fast}. An attractive description of the first mechanical scanning probe for ophthalmic echography based on a small piezoelectric ultrasound motor is reported in \cite{matani2006phase}. A novel method for line restoration in speckle images addressing a sparse estimation problem using both convex and non-convex optimization techniques is reported in \cite{anantrasirichai2017line}. Lung ultrasound imaging is a fast-evolving field of application for ultrasound technologies. The authors of \cite{demi2020real} design an image formation process to work on lung tissue, and ultrasound images are generated with four orthogonal bands centered at 3, 4, 5, and 6 MHz which can be acquired and displayed in real-time.
In the field of seismic tomography, several works have been performed. In \cite{berryman1989weighted} scientific methods are developed for the design of linear tomographic reconstruction algorithms with specified properties. An experimentation reconstruction method for seismic transmission travel-time tomography has been developed in \cite{song1999singular}. The method is implemented via the combinations of singular value decomposition, appropriate weighting matrices, and variable regularization parameters. An implementation of cross-correlation of environmental seismic noise from 1 month of recordings at USArray stations in California that produced hundreds of seismic wave group velocity measurements was done in \cite{shapiro2005high}. The authors used these measurements to construct tomographic images of the principal geological units of California, with low-speed anomalies corresponding to the main sedimentary basins and high-speed anomalies corresponding to the igneous cores of the major mountain ranges. Through the study of the properties of ambient seismic noise recorded around an 18-month of a time window, demonstrated in \cite{brenguier2008towards} that changes within the Piton de la Fournaise volcano can be monitored continuously by measuring very small relative seismic velocity perturbations on the order of 0.05\%. The ability to record volcanic building inflation in this way should improve the ability to predict eruptions, their intensity, and potential environmental impact
The foundations of diffraction tomography for offset vertical seismic profiling and well-to-well tomography are presented in \cite{devaney1984geophysical}. Computer simulations are used for underground vertical seismic profiling. The influence of quality in tomographic reconstructions obtained via filtered back-propagation algorithm is investigated in \cite{witten1986shallow}. To better understand the volcanic phenomena acting on Montserrat, a subset of the data, recorded by several land stations located from the southeast to the northwest line, has been analyzed in \cite{paulatto2010upper}. The resulting velocity model reveals the presence of high-velocity underground body movements at the volcanic edifices cores. The development and assessment of three novel muon tracing methods and two scattering angle projection methods for muon-tomography are provided in \cite{liu2018muon}. The reconstructed images showed an expected improvement in image quality when compared with conventional techniques. 3-D models of the P-wave velocity (Vp), the ratio of P- to S-wave velocity (Vs), $\left(\frac{Vp}{Vs}\right)$, and the P-wave quality factor (Qp) are determined \cite{sherburn2006three}. Vp and $\left(\frac{Vp}{Vs}\right)$ models were determined by jointly inverting P travel-times and S–P travel-time intervals, and a Qp model by inverting $t * \frac{t}{Qp}$ observations derived from modeling the velocity amplitude spectrum of P wave arrivals. The upper crustal isotropic and radial anisotropic structures of Jeju Island offshore Volcano have been imaged in \cite{lee2021upper} processing series ambient noise data from temporary seismic networks. Results of an ambient seismic noise tomography study of the Merapi–Merbabu complex are presented in \cite{yudistira2021imaging}. A seismic survey for the characterization of the main subsurface features of the Solfatara was developed in \cite{letort2012high}. Using the complete data set, the authors have carried out surface wave inversion with high spatial resolution. A classical minimization of a least-squares objective function was computed to retrieve the dispersion curves of the surface waves. The recognition and localization of magmatic fluids are pre-requisites for evaluating the volcano hazard of the highly urbanized area of Mt Vesuvius. Evidence and constraints for the volumetric estimation of magmatic fluids underneath this sleeping volcano have been studied in \cite{agostinetti2008seismic}. Experimental measurements revealed the presence of magma at relatively shallow depths. The volume of fluids (approx. 30 $km^3$) is sufficient to contribute to future explosive eruptions.
Finally controlled source audio-magnetotelluric (CSAMT) soundings tomography performed in the past in the volcanic area of Mt. Vesuvius by\cite{troiano2008shallow}, allowed a reliable electrical structure to be recovered down to a few km of depth.

In the field of vibration estimation by processing SAR images other relevant works are recalled. A new critical infrastructure monitoring procedure was developed in \cite{biondi2020monitoring}. The technique was applied to COSMO-SkyMed data, to detect and monitor the destabilization of the Mosul dam, which is the largest hydraulic structure in Iraq. The procedure consists of an in-depth modal evaluation based on micro-motion (m-m)  estimation through Doppler tracing of sub-apertures and multi-chromatic analysis (MCA).
On the other hand, a comprehensive damage detection procedure was designed using micro-motion estimation of critical sites based on modal property analysis developed in \cite{biondi2020perspectives}. Specifically, the m-m is processed to extract modal features such as natural frequencies and mode shapes generated by large infrastructure vibrations. Several case studies were considered, and the "Morandi" bridge (Polcevera Viaduct) in Genoa, Italy, was analyzed in-depth, highlighting anomalous vibration modes in the period before the bridge collapse.
After conducting a deep analysis of the state of the art, it becomes clear that to date there is a lack of a versatile, cost-effective, and widely applicable tool for internal monitoring of volcanoes. In this context, a new method that employs m-m estimation to perform deep scanning of volcanoes is presented. This approach appears to be tailored to fill this gap-technology when today there is direct access to SAR satellite data, which is a day-night sensor, and also able to penetrate clouds. 
The experimental results are obtained by processing one SAR Spotlight image, observing the Vesuvius, and revealing its internal structure. Tomographic maps show the main crater section, the volcanic conduit, and the main mouth. In-depth, the magma chamber and the presence of a secondary lava tube are also visible. In addition, the internal structure of the currently plugged main lava tube, reveals several different structures of soil layers.

\section{Methodology}
\label{Methodology}

In this work, the m-m technique is used to perform sonic imaging by processing a single synthetic aperture radar (SAR) image in the single-look-complex (SLC) configuration. The technique involves the m-m estimation belonging to the Vesuvius volcano and is generated by the intensive underground seismic activity that reflects superficial vibrations. The m-m estimation is done through MCA, performed in the Doppler direction. Multiple Doppler sub-apertures, SAR images with lower azimuth resolution, are generated to estimate the vibrational trend of some pixels of interest. The infra-chromatic displacement is calculated through the pixel tracking technique \cite{filippo2019cosmo,biondi2019micro}, using highly performance sub-pixel coregistration \cite{biondi2020monitoring,biondi2020perspectives}. Vibrations observed along the tomographic view-direction, embedded into the multi-chromatic Doppler diversity, are focused along the height (or depth) dimension, and developing high-resolution tomographic underground imaging. 
	
\begin{figure}[htp]
    \centering
    \includegraphics[width=12.0cm,height=9.0cm]{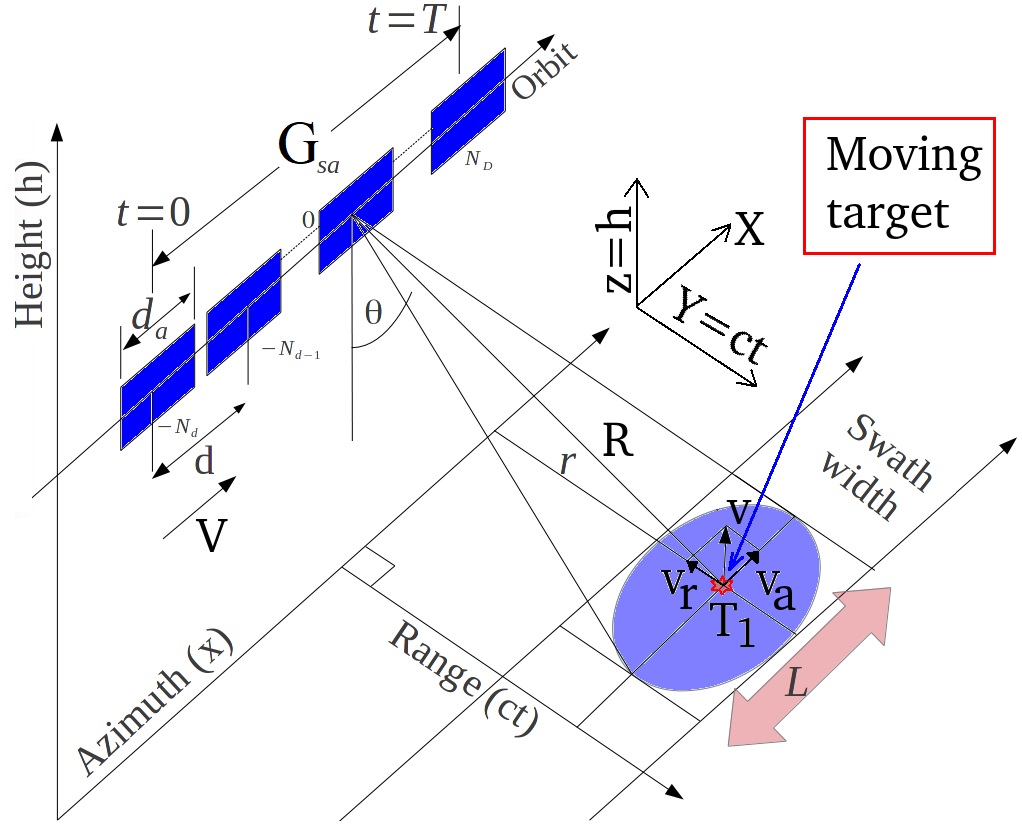}
    \caption{SAR acquisition geometry.}
    \label{Geometry_1}
\end{figure}

The SAR synthesizes the electromagnetic image through a ''side looking'' acquisition, according to the observation geometry shown in Figure \ref{Geometry_1}, where:
\begin{itemize}
\item $r$ is the zero-Doppler distance (constant); 
\item $R$ is the slant-range;
\item $R_0$ is the reference range at $t=0$;
\item $d_a$ is the physical antenna aperture length;
\item $V$ is the platform velocity;
\item $d$ is the distance between two range acquisitions;
\item $G_{sa}$ is the total synthetic aperture length;
\item $t$ is the acquisition time variable;
\item $T$ is the observation duration;
\item $t=0$ and $t=T$ are the start and stop time acquisition respectively;
\item $L=\frac{\lambda r}{d_a}$ is the azimuth electromagnetic footprint width;
\item $\theta$ is the incidence angle of the electromagnetic radiation pattern.
\end{itemize}
All the above parameters are related to the staring-spotlight SAR acquisition that is adopted in this work.
Considering what is formalised, the MCA technique, based on Doppler sub-apertures, is used to estimate the following master-slave pixel shift complex parameters:
\begin{itemize}
\item $\epsilon_{r_1}=\frac{v_r}{V}$ (due to range velocity);
\item $\epsilon_{r_2}=\frac{a_rR_0}{V^2}$ (due to range acceleration);
\item $\epsilon_{c_1}=\frac{v_c}{V}$ (due to azimuth velocity).  
\end{itemize}
Thus, the above terms modify the received signal, as shown in \cite{raney1971synthetic}, and should be taken into account in Equation.\

\subsection{Headings: second level}
\lipsum[5]
\begin{equation}
\xi _{ij}(t)=P(x_{t}=i,x_{t+1}=j|y,v,w;\theta)= {\frac {\alpha _{i}(t)a^{w_t}_{ij}\beta _{j}(t+1)b^{v_{t+1}}_{j}(y_{t+1})}{\sum _{i=1}^{N} \sum _{j=1}^{N} \alpha _{i}(t)a^{w_t}_{ij}\beta _{j}(t+1)b^{v_{t+1}}_{j}(y_{t+1})}}
\end{equation}

\begin{figure}[htp]
    \centering
    \includegraphics[width=15.0cm,height=4.0cm]{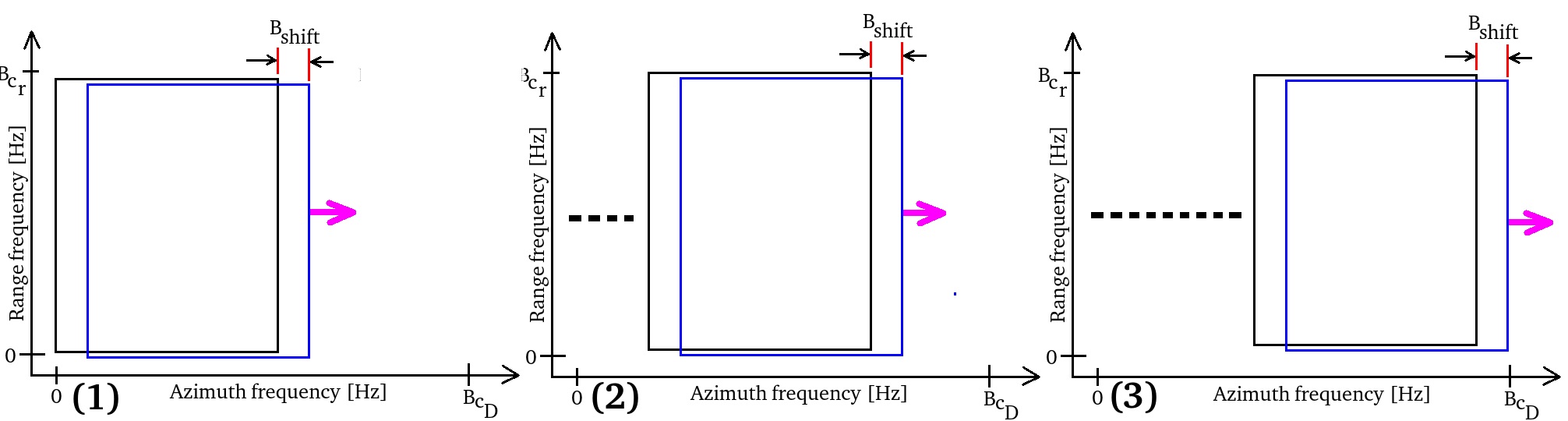}
    \caption{Doppler sub-aperture strategy.}
    \label{Bandwidth_Strategy_1}
\end{figure}

\subsection{Tomographic Model}\label{Methods}
\begin{figure}[htp]
    \centering
    \includegraphics[width=9.0cm,height=10.0cm]{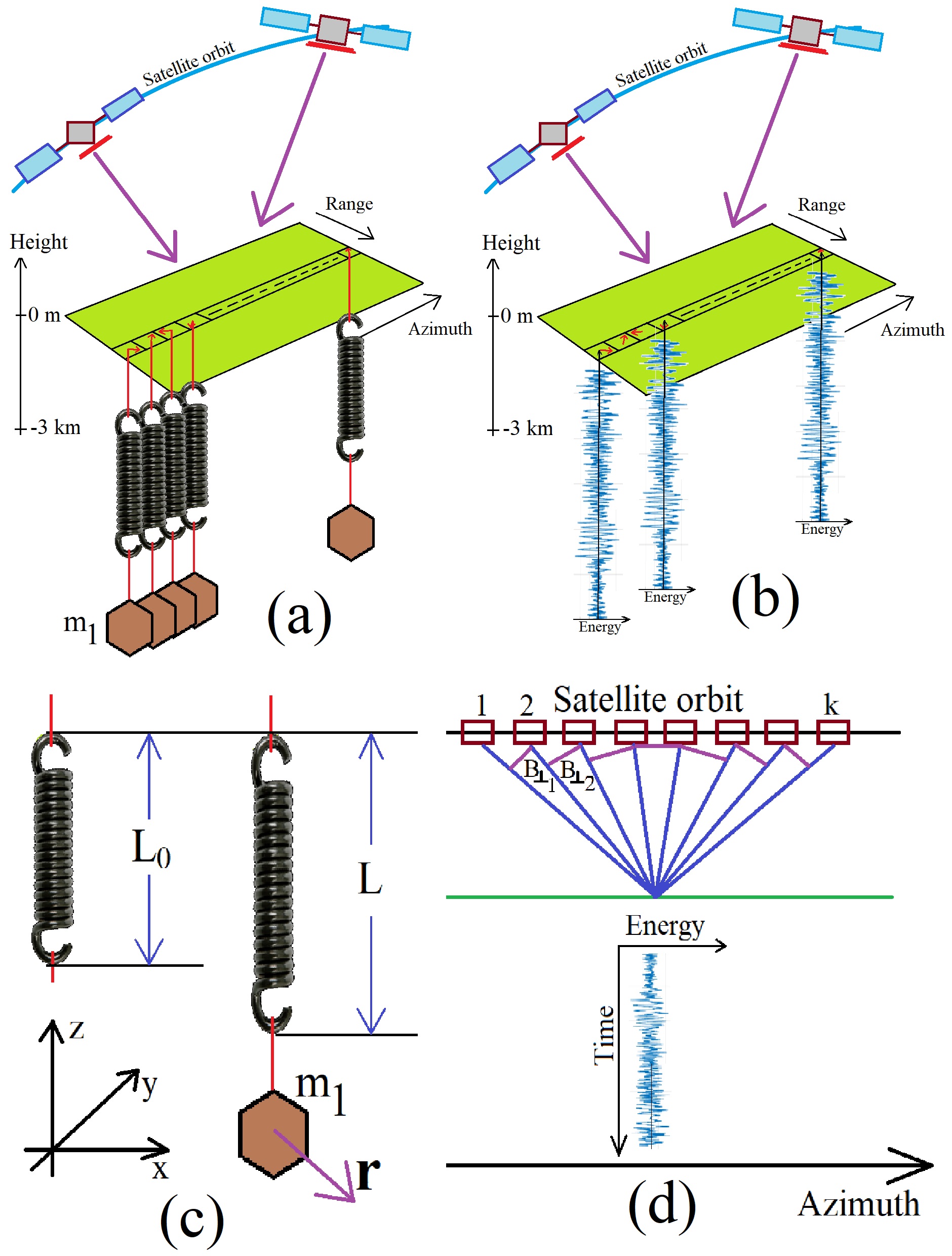}
    \caption{Tomographic acquisition geometry.}
    \label{Tomo_1}
\end{figure}

Considering a single SLC image from which we applied the MCA according to the frequency allocation strategy depicted in Figure \ref{Bandwidth_Strategy_1}, the tomogram represented by the line of contiguous pixels shown in Figure \ref{Tomo_1} is calculated. The vibrations present on the tomographic plane extending from the earth's surface to a depth of a few kilometres is assessed. The figure represents a series of harmonic oscillators anchored on each pixel of the tomographic line, symbolically represented as a spring linked to a mass and oscillating due to the application of harmonic vibrations. Each wave generated by each harmonic oscillator bounces off the surface of the earth as there is an abrupt variation in the density of the medium (the ground-air boundary). On each pixel a vibrational phasor is observed in time applying Doppler MCA \cite{biondi2020monitoring,biondi2020perspectives}. 
Through the orbital change of view (which is performed in azimuth), an effective subsurface in-depth vibrational scan of the Earth is achieved.  
	
\subsection{Vibrational Model of the Earth}\label{Vibrational_Model_of_cables}
The proposed vibrational model of the earth surface is schematically shown in Figure \ref{Tomo_1} (a), and (b). The geometrical reference system for both sub-pictures is the range, azimuth and altitude three-dimensional space. For the present case, the vertical dimension represents the depth below the topographic level, (for this specific case the medium boundary is represented by the green plane). The tomographic line of interest is constituted of the series of contiguous pixels laying on the green plane. As can be seen from Figure \ref{Tomo_1} (a) on each pixel belonging to the tomographic line, a mass is hanging using a spring. This system is now induced to oscillate harmonically, helped by the earth magma instability. These oscillations are schematized as the vibration energy function visible in Figure \ref{Tomo_1} (b). In this context the radar instantaneously perceives this coherent harmonic oscillation. In a mathematical point of view, the earth displacement is perceived as a complex shift belonging on each pixel of interest. Each instantaneously displacement is estimated between the master image with respect to the slave, where oversimplification shifts are estimated through the pixel tracking technique \cite{biondi2020monitoring,biondi2020perspectives}. The number of tomographic independent looks (depending on the total number of Doppler sub-apertures) are defined by the parameter $k$.

We suppose now the spring being perturbed by an impulsed force. According to this perturbation the rope begins to vibrate describing an harmonic motion (in this context we are not considering any form of friction). Resulting perturbation moves the rope through the space-time in the form of a sinusoidal function. The seismic wave will then reach a constraint end that will cause it to reflect in the opposite direction. The reflected wave will then reach the opposite constraint that will make it reflect in the original direction and returning in the initial location, maintaining the same frequency and amplitude. According to Classical Physics principles, the rebounding wave is superimposed on the arriving wave, and the interference of two sine waves with the same amplitude and frequency propagating in opposite directions leads to the generation of an ideal and perpetual standing wave on the spring. Each vibrational channel is now considered when the spring is able to oscillate into the three-dimensional space, according to specific perturbation nature. When the earth vibrates, it happens that the length of the spring must also fluctuate. This phenomenon causes oscillations in the tension domain of the spring. It is clear that these oscillations (i.e. the longitudinal ones) propagate through a frequency approximately twice as high as the frequency value of the transverse vibrations. The coupling between the transverse and longitudinal oscillations of the spring can essentially be modeled through non-linear phenomena.

\begin{figure}[htp]
    \centering
    \includegraphics[width=14.0cm,height=6.5cm]{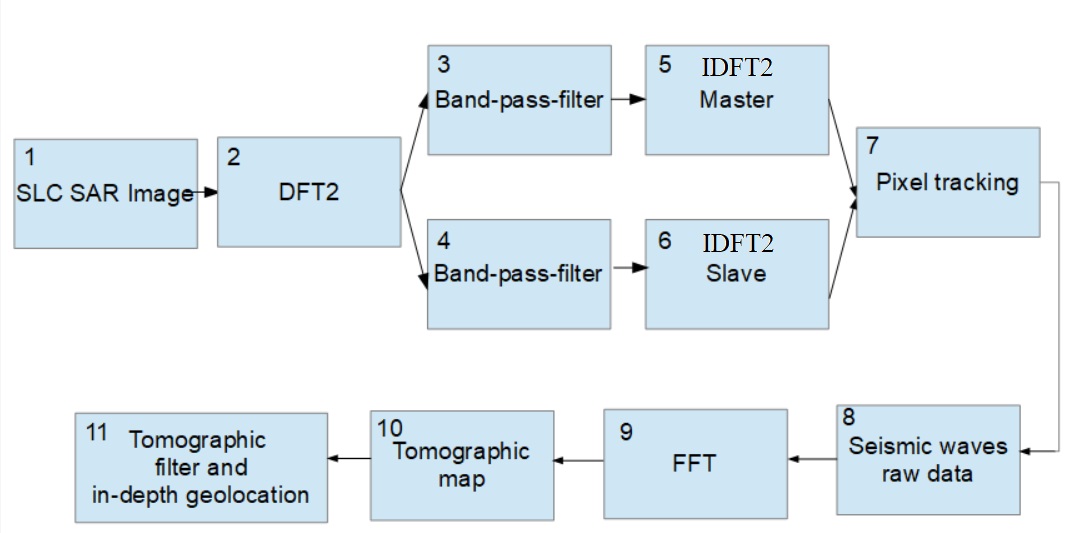}
    \caption{Computational scheme of tomographic imaging.}
    \label{Computational_scheme_1}
\end{figure}

Figure \ref{Tomo_1} (c, d) illustrates the oscillating model in the Euclidean space-time coordinates (x,y,z,t), where the satellite motion has been purified from any orbital distortions, so that the geometric parameters used to perform the tomographic focusing can be rigorously understood. From Figure \ref{Tomo_1} (c), $L$ is the length of the spring when it is at its maximum tension while $L_0$ is its length when no mass is present. Finally the spring has been considered to have an elastic constant equal to $\xi$. The vibrational force applied to the mass $m1$ of Figure \ref{Tomo_1} (c) is equal to \cite{tufillaro1989nonlinear}:

$F=-4 \xi \mathbf{r} \left(1-\frac{L_0}{\sqrt{L^2+4\mathbf{r}^2}}\right)$.\label{Eq_16}

If $\mathbf{r}\ll L$, \ref{Eq_16} is expanded in the following series:

$F=-4 \xi \mathbf{r}(L-L_0)\left(\frac{\mathbf{r}}{L}\right)-8 \xi L_0 \left[\left(\frac{\mathbf{r}}{L}\right)^3-\left(\frac{\mathbf{r}}{L}\right)^5+\dots\right]$\label{Eq_16_Bis},

where a precise approximation of \ref{Eq_16_Bis} is the following cubic restoring force:

\label{Eq_18}
$F=m\ddot{\mathbf{r}}\approx-4 \xi \mathbf{r}(L-L_0)\left(\frac{\mathbf{r}}{L}\right) \left[1+\frac{2L_0}{(L-L_0)}\left(\frac{\mathbf{r}}{L}\right)^2\right]$.

Considering \ref{Eq_18}, the non-linearity dominates when $L \approx L_0$.
	If we define:
	\label{Eq_19}
	$\omega_0=\frac{4 \xi}{m}\left[\frac{\left(L-L_0\right)}{L}\right]$,
	and
\label{Eq_20}
$\xi=\frac{2L_0}{L^2}\left(L-L_0\right)$.

Considering \ref{Eq_18} we have:

\label{Eq_21}

$\ddot{\mathbf{r}}+\omega^2 \mathbf{r} \left(1+ \xi\mathbf{r}^2\right)=0$.
If we consider damping and forcing \ref{Eq_21} is modified as:
\label{Eq_24}
$\ddot{\mathbf{r}}+\lambda\dot{\mathbf{r}}+\omega^2\left(1+\xi\mathbf{r}^2\right) \mathbf{r}=\mathbf{f}(\omega t)$,

where $\mathbf{f}(\omega t)$ is the forcing term and $\lambda$ is the damping coefficient. 
If non-linearity of \ref{Eq_24} is sufficiently low, it can be reduced into the following two-degree-of-freedom linear harmonic oscillator:

\label{Eq_23}
$\mathbf{r}(t)=\left(a \cos\omega_0 t, b\sin \omega_0 t\right)\exp\left(\frac{-\lambda t}{2}\right)$.

In \ref{Eq_23} $\{a,b\}$ are the instantaneous shifts estimated by the coregistrator. The harmonic oscillator ... is the displacement parameters ${\epsilon_{r_1}, \epsilon_{r_2}, \epsilon_{c_1}}$ estimated by .
According to Figure \ref{Tomo_1} (d) the vector representation of $k$ samples of the time-domain function \ref{Eq_23} consisting in the following multi-frequency data input is considered:

	\label{eq_2}
	$\mathbf{Y} =\left[\mathbf{y}(1),\dots ,\mathbf{y}(k)\right],\in \mathbf{C}^{k\times 1}$.
	
	The steering matrix $\mathbf{A}(z)=\left[\mathbf{a}(z_{min}),\dots,\mathbf{a}(z_{MAX})\right]$, $\in  	\mathbf{C}^{k\times F}$ contains the phase information of to the Doppler frequency variation of the sub-aperture strategy, associated to a source located at the elevation position $\mathbf{z} \in \{z_{min},z_{MAX}\}$,
	
	\label{Eq_31}
		$\mathbf{A}(\mathbf{K}_z,\mathbf{z})=$
			$1,\exp(\jmath 2\pi k_{z_2} t z_0),\dots,\exp(\jmath 2\pi k_{z_{k-1}} t z_0)$ \\
			$1,\exp(\jmath 2\pi k_{z_2} t z_1),\dots,\exp(\jmath 2\pi k_{z_{k-1}} t z_1)$ \\
			$\dots$ \\
			$1,\exp(\jmath 2\pi k_{z_2} t z_{F-1}),\dots,\exp(\jmath 2\pi k_{z_{k-1}} t z_{F-1})$,
	
	where $\mathbf{K}_z= \frac{4\pi B_{\perp}}{\lambda \mathbf{r}_i \sin \theta}$, $i=1,\dots,k$, $B_\perp$ is the $i-$th orthogonal baseline which is visible in Figure \ref{Tomo_1} (d), and $\mathbf{r}_i$ is the $i-$th slant-range distance. The standard sonic tomographic model is given by the following relation:
	
	\label{Eq_35}
	$\mathbf{Y}=\mathbf{A}(\mathbf{K}_z,\mathbf{z}) {\mathbf{h}}(\mathbf{z})$.
	
	where in \ref{Eq_35} ${\mathbf{h}}(\mathbf{z}) \in \mathbf{C}^{1 \times F}$, inverting \ref{Eq_35} I finally find the following tomographic solution:
	
	\label{Eq_34}
	${\mathbf{h}}(\mathbf{z})=\mathbf{A}(\mathbf{K}_z,\mathbf{z})^\dagger \mathbf{Y}$.
	
In the \ref{Eq_34} the steering matrix $\mathbf{A}(\mathbf{K}_z,\mathbf{z})$ represents the best approximation of a matrix operator performing the digital Fourier transform (DFT) of $\mathbf{Y}$. The tomographic image $\mathbf{h}(\mathbf{z})$, which represents the spectrum of $\mathbf{A}(\mathbf{K}_z,\mathbf{z})$, is obtained by doing pulse compression.
	
The tomographic resolution is equal to $\delta_T=\frac{\lambda R}{2 A}$, where $\lambda$ is the sound wavelength over the earth, $R$ is the slant range, and $A$ is the orbit aperture considered in the tomographic synthesis, in other words, consists to the Doppler bandwidth used to synthesize the sub-apertures. The maximum tomographic resolution obtainable using this SLC data, synthesized at 24 kHz, is as follows. Considering an average speed of propagation of the seismic waves of about $v \approx 3500 \frac{k m}{h}$ (approximately $972 \frac {m}{s}$), a frequency of investigation set by us equal to 200 Hz, the wavelength of these vibrations is equal to about $\lambda=\frac{v}{f}\approx \frac{3500}{200} \approx 4.86 m$. Considering the above parameters, extending the tomography to the maximum orbital aperture equal to half the total length of the orbit, therefore about $42000$ m, with $R=650000 m$ the tomographic resolution is equal to $\delta_z=\frac{\lambda R}{2 A}= \frac{4.86 \cdot 650000}{2 \cdot 42000} \approx 36 m$. This is the tomographic resolution set to calculate all the experimental parts shown in section \ref{Experimental Results}.

\section{Experimental Results}\label{Experimental Results}
\begin{figure}[htp]
    \centering
    \includegraphics[width=13.0cm,height=7.0cm]{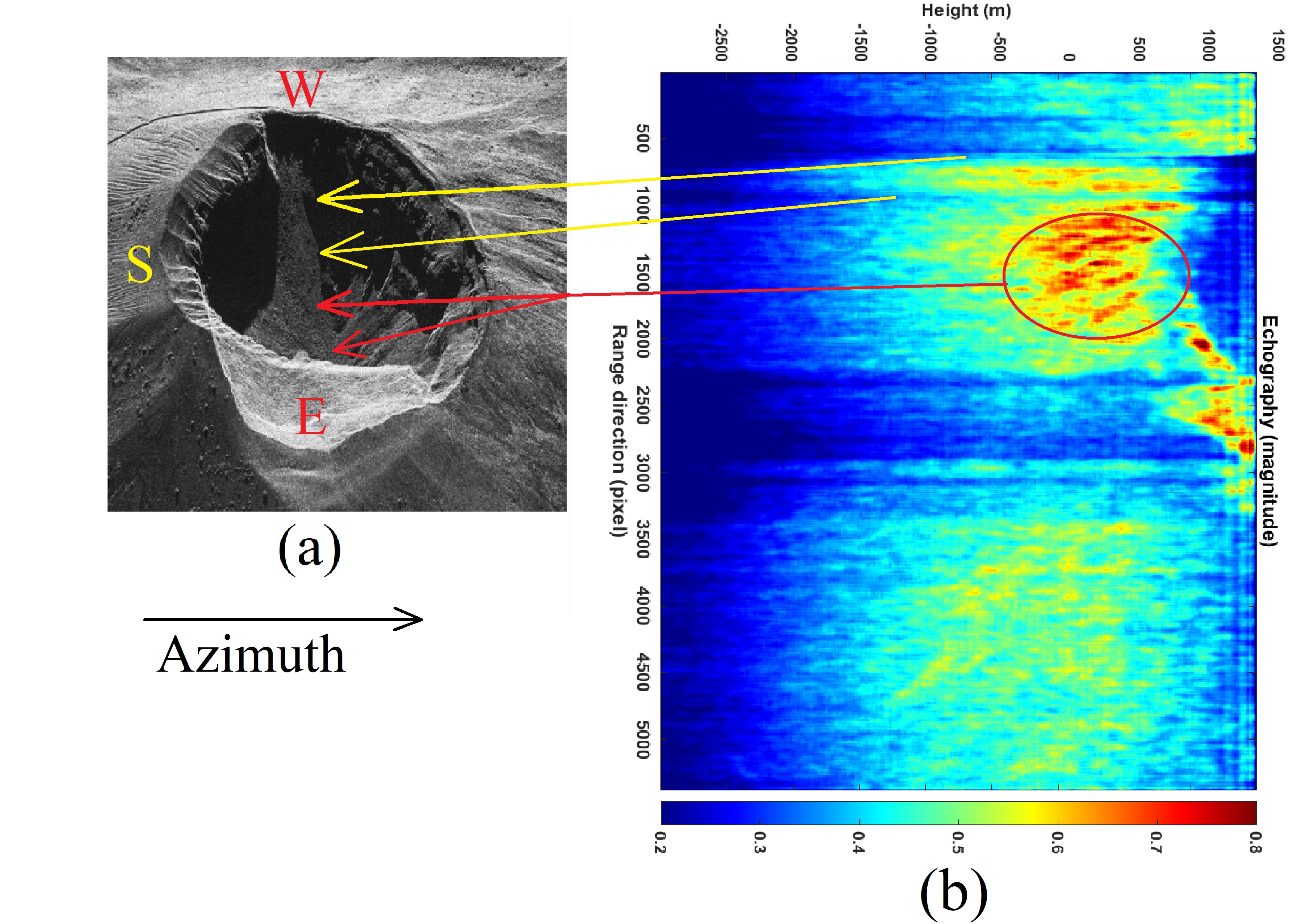}
    \caption{Range-line echographic tomography detailed result (magnitude). (a): SLC SAR image (in magnitude). (b): Tomographic result (in magnitude).}
    \label{Vesuvio_Particular_6}
\end{figure}

\begin{figure}[htp]
    \centering
    \includegraphics[width=12.0cm,height=4.0cm]{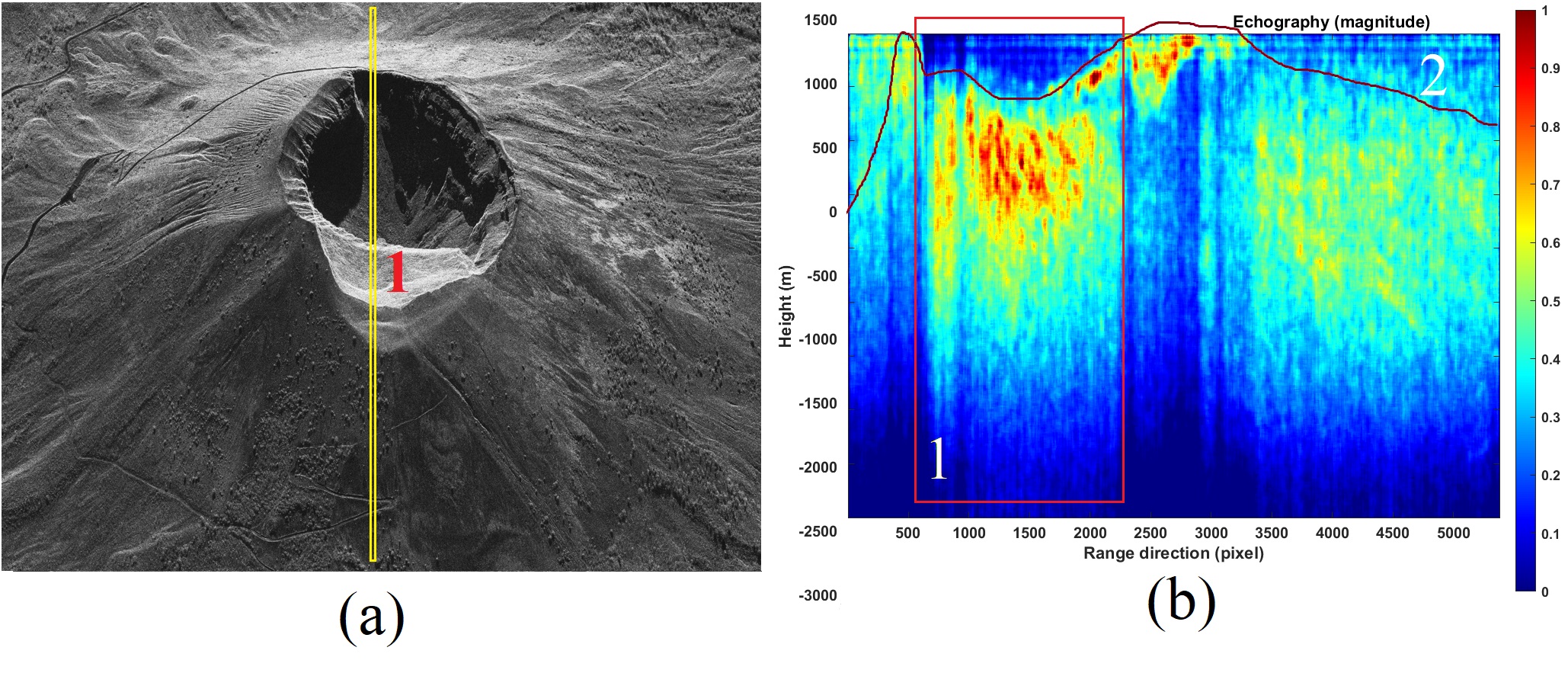}
    \caption{Range-line echographic tomography result (magnitude). (a): SLC SAR image (in magnitude). (b): Tomographic result (in magnitude).}
    \label{Range_599_1}
\end{figure}

In this section, all the experimental results are described by processing a SAR image acquired by the COSMO-SkyMed Second Generation (CSG) satellite constellation. The data concerns a spotlight-2A acquisition-mode, in the horizontal-horizontal (HH) polarization, with a Doppler band of about 22.5 kHz, and a chirp band of about 450 MHz. For this first case study, both the optical and the SLC image in magnitude are depicted in Figures (a), and (b) respectively. The characteristics of the employed SAR image are listed in Table. Figure \ref{Range_599_1} (a) is the representation of the SAR (magnitude) image where Vesuvius with its main crater is visible. Yellow line 1 is where the seismic tomogram is calculated. The test data-set is constituted by an entire range line belonging between the near-range and the far-range of the SAR image. Figure \ref{Range_599_1} (b) is the result representing the estimated seismic tomography. The depth of investigation is about 3 km. In the figure, the surface topography of the volcano and its entire interior are visible. From the tomogram, it is possible to observe the crater of the main conduit inside the red box 1. From this result, we observe the presence of vibration energy accumulations located in depth. This discontinuity suggests the presence of material that vibrates more concerning the background. This anomaly could be associated with the presence of denser material, and therefore with a sort of cap located below the main crater of the Vesuvius. On the left side of the cap, there are two ducts, probably created by the pressure of the vapors or smoke coming from underground. The maximum depth measured by the tomgram is approximately 3 km from the maximum topographic surface height of the volcano. Figure \ref{Vesuvio_Particular_6} (a), (b) is a detailed analysis of the of the volcano lava tube. In this figure, two apertures, probably excavated by high-pressure lava gases, and the presence of a large portion of a denser material (having higher vibrational energy), residing below the surface of the main crater at a depth of about 1 km, are visible.

\begin{figure}[htp]
    \centering
    \includegraphics[width=12.0cm,height=4.0cm]{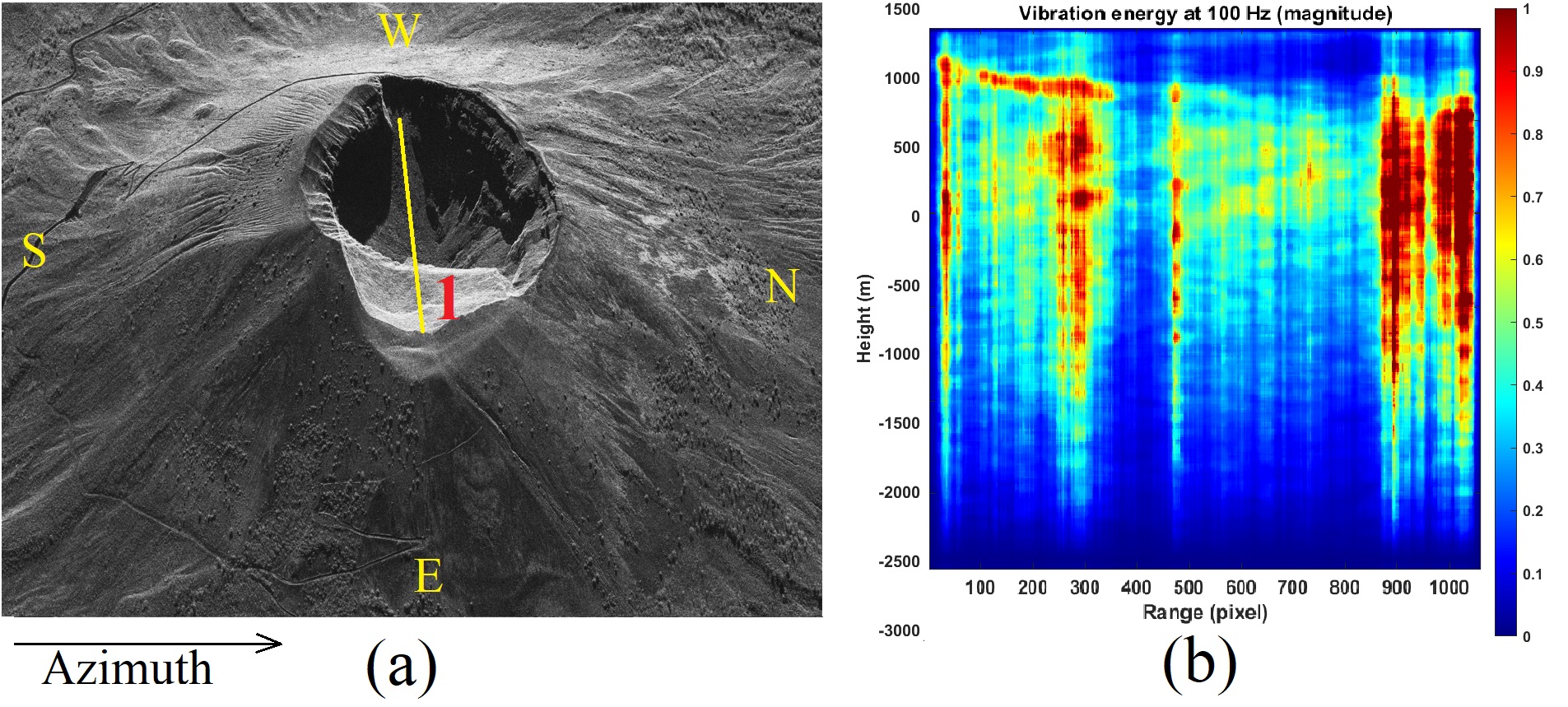}
    \caption{Range-line echographic tomography result (magnitude). (a): SLC SAR image (in magnitude). (b): Tomographic result (in magnitude).}
    \label{Range_599_3}
\end{figure}

\begin{figure}[htp]
    \centering
    \includegraphics[width=12.0cm,height=4.0cm]{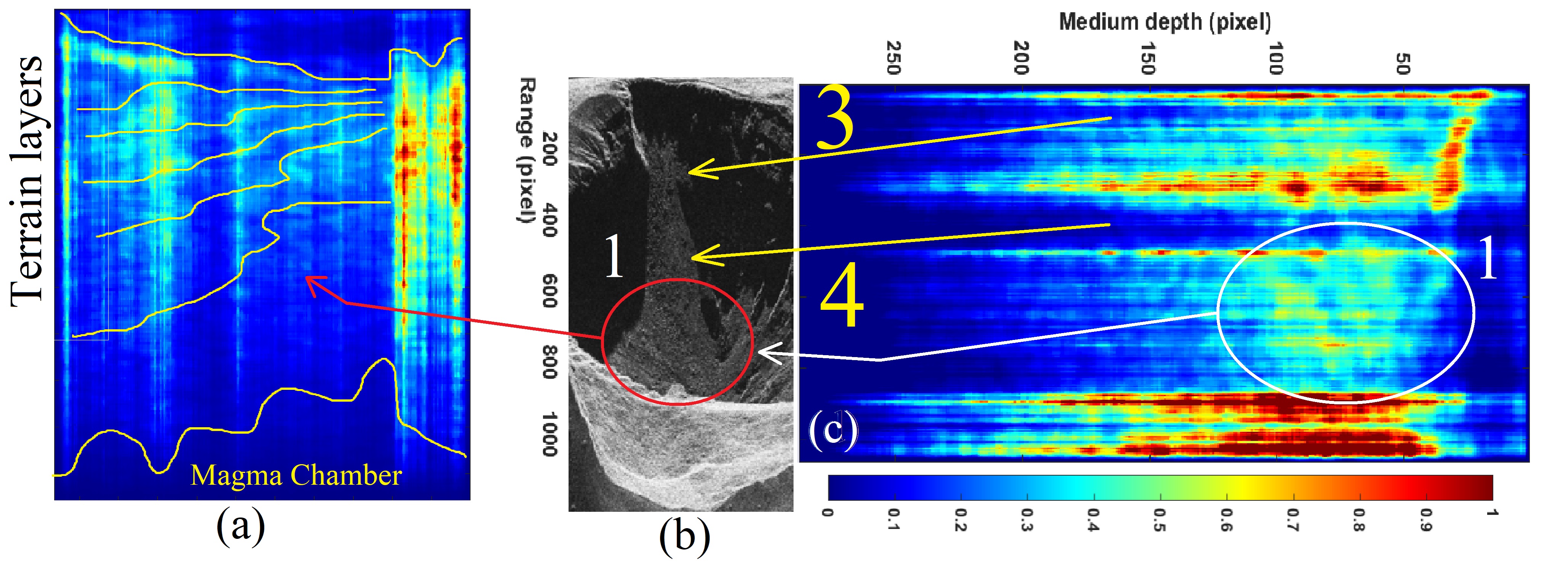}
    \caption{Range-line echographic tomography result (magnitude). (a): Tomographic result (in magnitude). (b): SLC SAR image (in magnitude). (c): Tomographic result (in magnitude).}
    \label{RR1}
\end{figure}

\begin{figure}[htp]
    \centering
    \includegraphics[width=12.0cm,height=5.0cm]{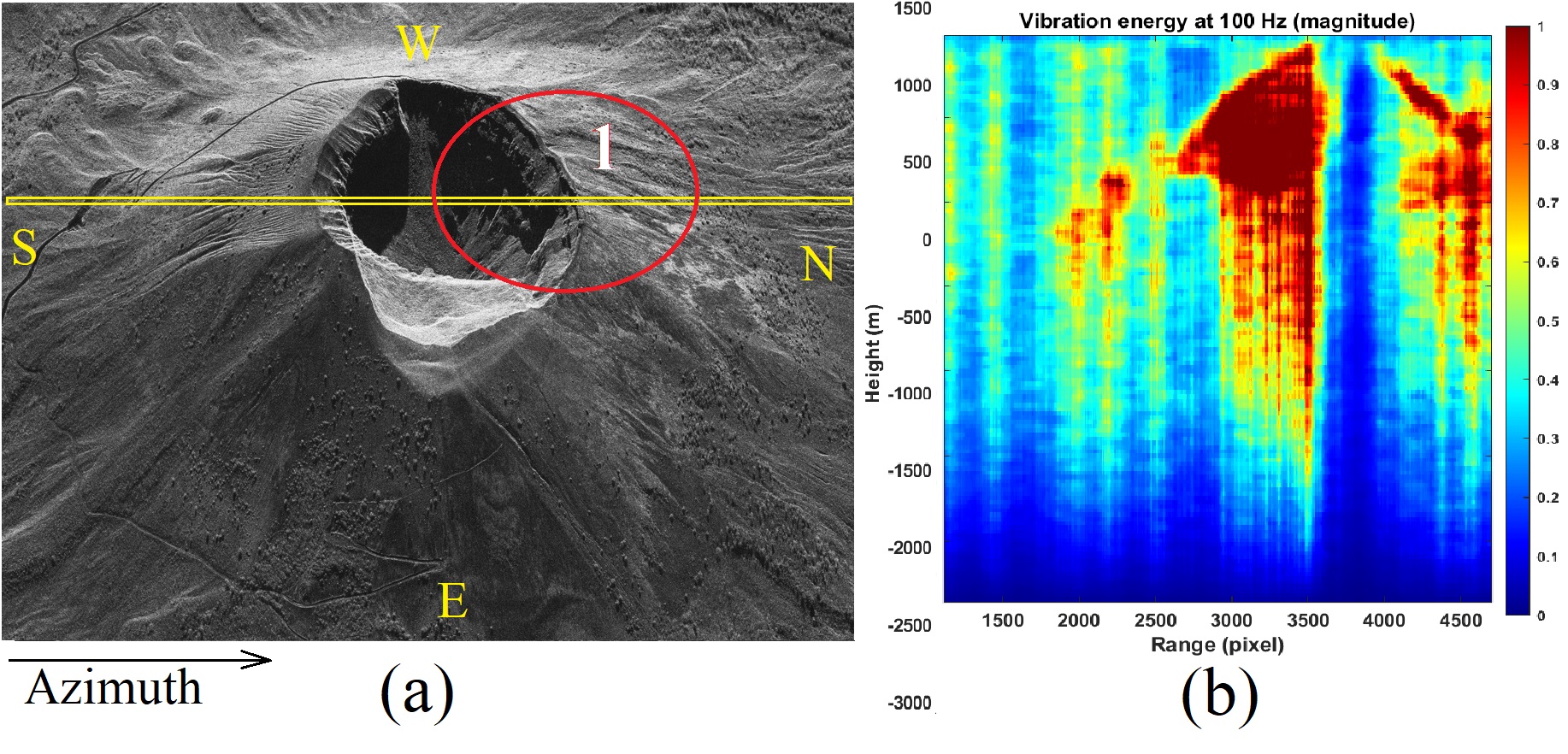}
    \caption{Range-line echographic tomography result (magnitude). (a): SLC SAR image (in magnitude). (b): Tomographic result (in magnitude).}
    \label{Range_599_4}
\end{figure}

\begin{figure}[htp]
    \centering
    \includegraphics[width=12.0cm,height=8.0cm]{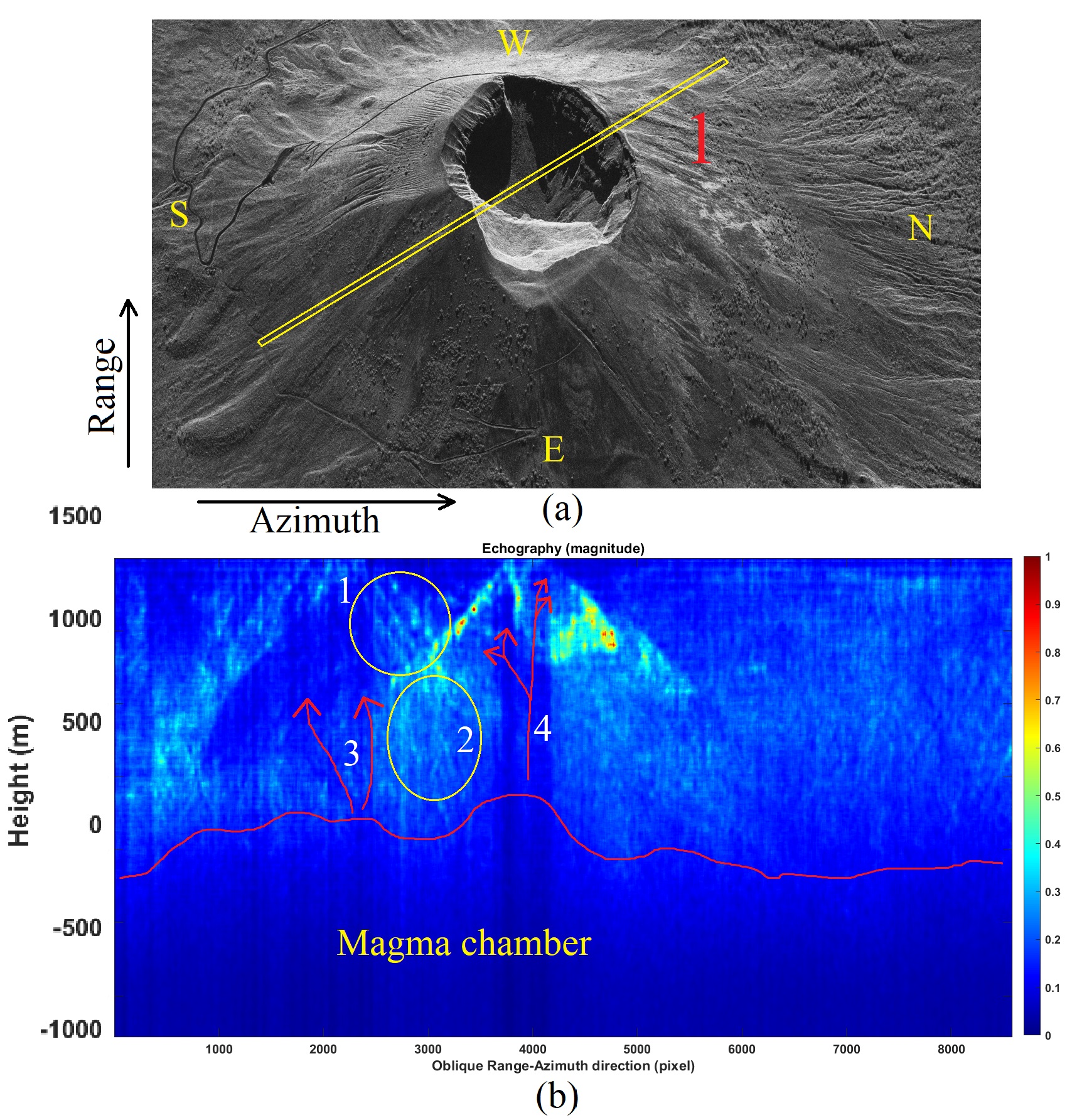}
    \caption{Range-azimuth line echographic tomography result (magnitude). (a): SLC SAR image (in magnitude). (b): Tomographic result (in magnitude).}
    \label{Obliqua_1}
\end{figure}

Figure \ref{Range_599_3} (a) shows the SAR image (in magnitude) of the Vesuvius according to a new tomographic line inserted, slightly more inclined, to cover as much as possible, in radar visibility, the volcano crater diameter, in the condition of having the greater scattering energy. On those pixels, we calculated the seismic tomogram visible in Figure \ref{Range_599_3} (b). In this case, the possible presence of a cap constituted by dense rocky material and the presence of an energetic gap where the seismic waves oscillate with less intensity is visible. In Figure \ref{RR1} (a) the detailed seismic tomography of the Vesuvius crater is depicted, and details of the SAR image are proposed in Figure \ref{RR1} (b). Figure \ref{RR1} (c) is the sonic tomogram of a range line spanning the entire crater. Through yellow arrows 3 and 4, pointing to the SAR image, two gas conduits are shown, finally, through arrow 4, the presence of a massive soil accumulation is detected. This facility appears to be composed of several layers having different physical compositions. The soil layers are detailed within circle 1 (visible in Figure \ref{RR1} (b)). The maximum depth of the layers is measured to be about 2 km deep, relative to the top surface of the volcano.
	Figure \ref{Range_599_4} is an overview of Vesuvius for which the seismic tomogram has been calculated through a purely azimuth-oriented line, which is visible in Figure \ref{Range_599_4} (a), and tracked through the yellow line 1. The tomographic result is shown in Figure \ref{Range_599_4} (b), for which the main section is shown within the red circle visible in Figure \ref{Range_599_4} (a). The result shows probable lava conduits, detected in terms of vibrational energy discontinuities. Figure \ref{Obliqua_1} represents the tomographic result obtained considering an oblique section, visible in Figure \ref{Obliqua_1} (a). For this case, the tomographic line is traced through yellow line 1. The proposed tomographic map shows the entire volcano internal section (Figure \ref{Obliqua_1} (b)). The tomogram was calculated at a higher resolution and taking into account a longer orbital variation.
	
\begin{figure}[htp]
    \centering
    \includegraphics[width=12.0cm,height=7cm]{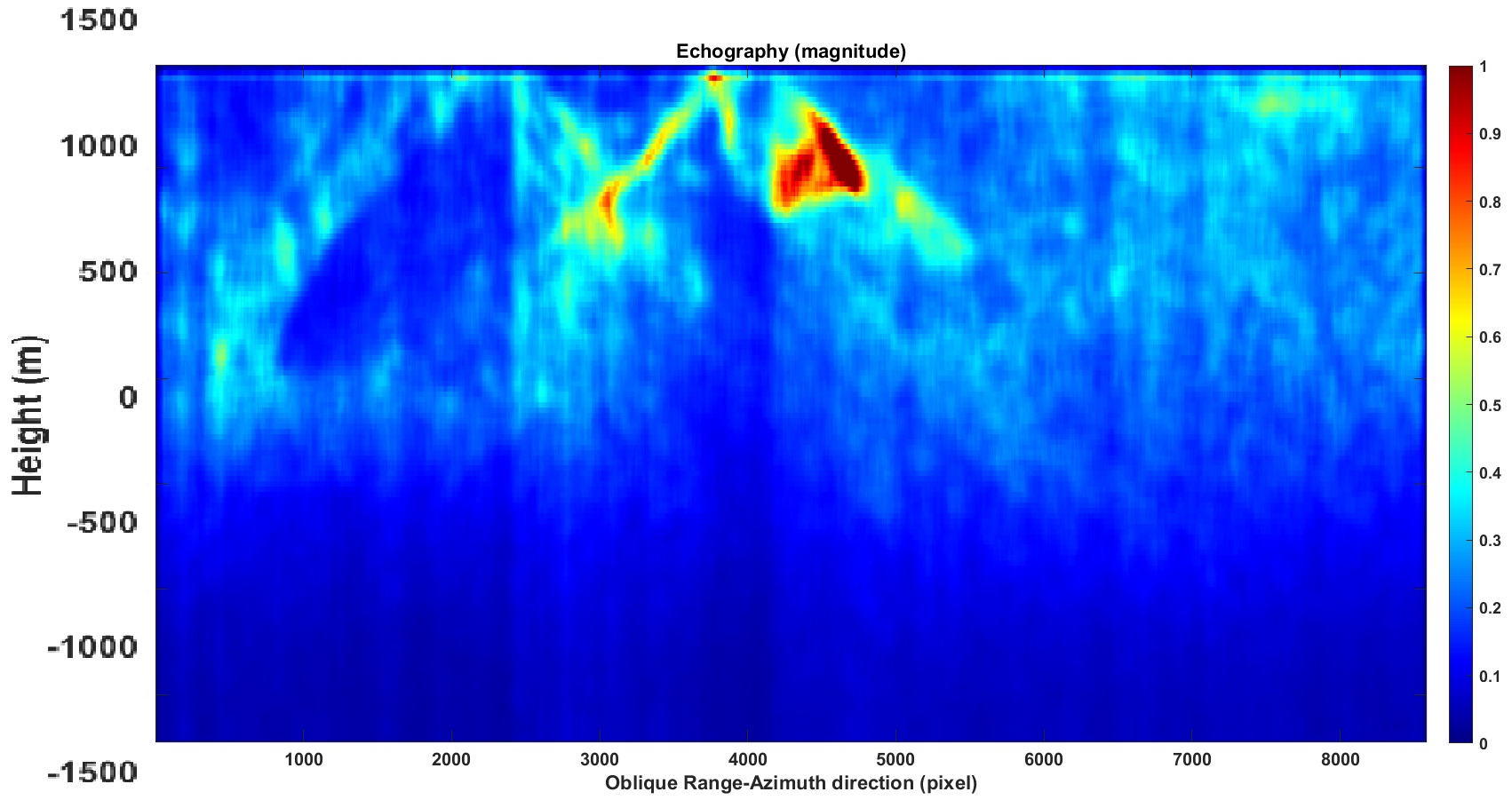}
    \caption{Range-azimuth line echographic tomography result (magnitude).}
    \label{Obliqua_2}
\end{figure}

	The following information is obtained from the calculated tomograms:
	\begin{itemize}
		\item The inner western volcano crater tomographic slopes consist of layered material, visible within the yellow circle 1 of Figure \ref{RR1} (b);
		\item The main volcano crater can be plugged by a cap of material consistent with vibrations energy singularity, depicted within the yellow circle 2 of Figure \ref{Obliqua_1} (b);
		\item The main conduits from the magma chamber, rise from the east and west tomographic sections, indicated by arrows 3 and 4 of Figure \ref{Obliqua_1} (b);
		\item the same results are obtained by calculating the sonic tomogram at a lower vibrational frequency, for which the result is visible in Figure \ref{Obliqua_2}.
	\end{itemize}
	
\begin{figure}[htp]
    \centering
    \includegraphics[width=16.5cm,height=13.5cm]{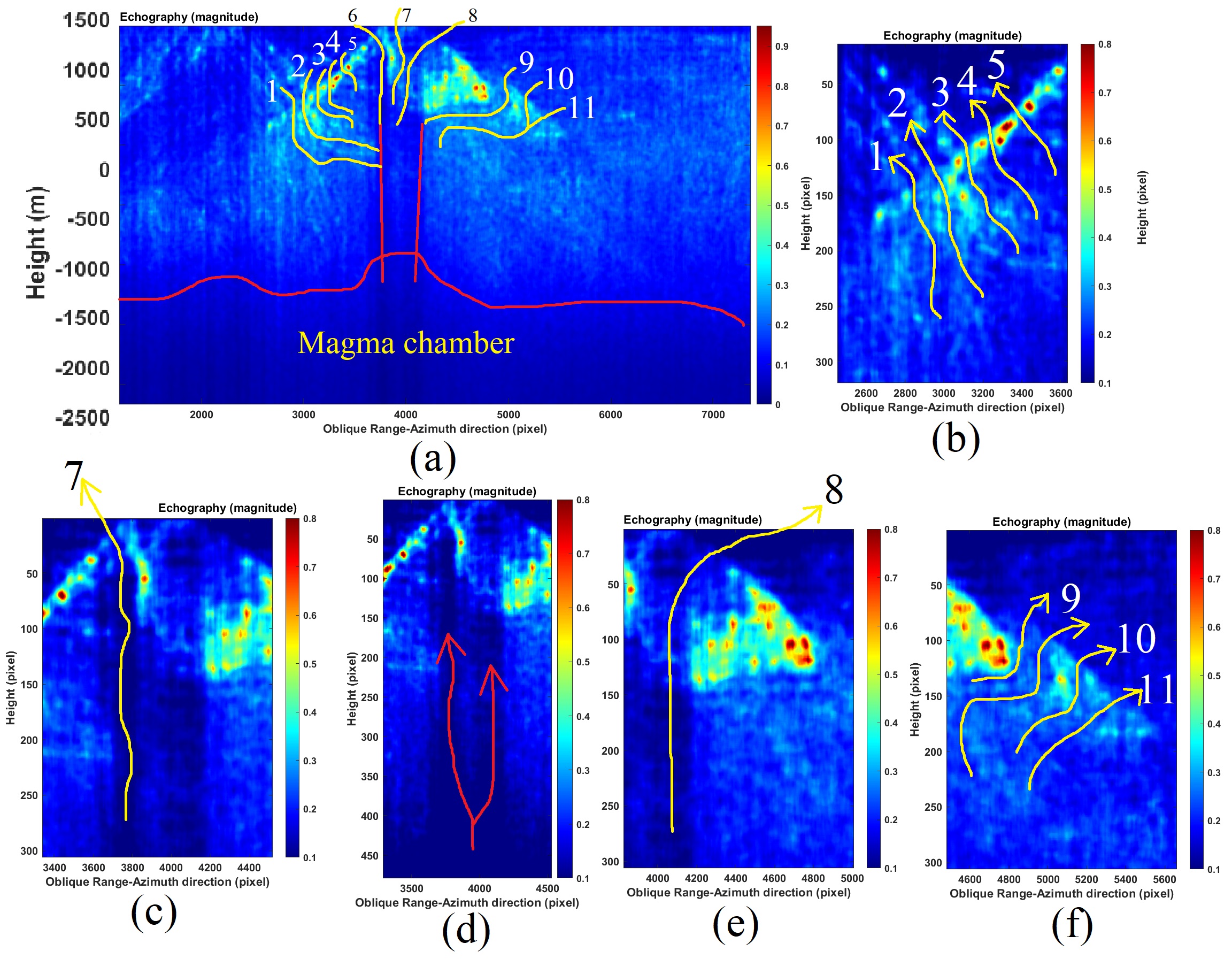}
    \caption{Tomographic results and analysis. (a): Tomographic analysis of the external eastern slope of Vesuvius, where potentially 11 lava apertures are visible. At the bottom of the tomogram, the presence of a magma chamber is evident. (b): Tomographic results and analysis. (a): Tomographic analysis of the internal eastern slope of Vesuvius, where potentially 5 lava apertures are visible. At the bottom of the tomogram, the presence of a magma chamber is evident. (c): Detail of the eastern slope showing an internal lava conduit. (d): Detail of the deepest part of the conduit showing a splitting of the lava tube. (e): Detail of the external lava tube (upper part). (f): Detail of the eastern slope on the external side (lower part).}
    \label{DDD}
\end{figure}

	A detailed list of vent conduits is proposed in Figure \ref{DDD}. These natural passages are made of material that does not resonate. We consider this characteristic as possible lava paths that a hypothetical new eruption might prefer. From the western side of the great crater, it is possible to edit 11 apertures. The ones belonging from 1 through 5 are located within the main crater, while apertures 9, 10, and 11 may extend outside the crater. The complete list of possible lava paths can be seen in Figure \ref{DDD} (a). A detailed tomographic section of the apertures finishing inside the crater can be seen in Figure \ref{DDD} (b), while a detailed tomogram of the upper lava ducts are visible in Figure \ref{DDD} (c), (d), and (e), and finally some external lava-paths are depicted in Figure \ref{DDD} (f).	

\section{Validation of the Sonic Tomographic SAR Results}
\subsection{Topographic Validation}
\begin{figure}[htp]
    \centering
    \includegraphics[width=12cm,height=9cm]{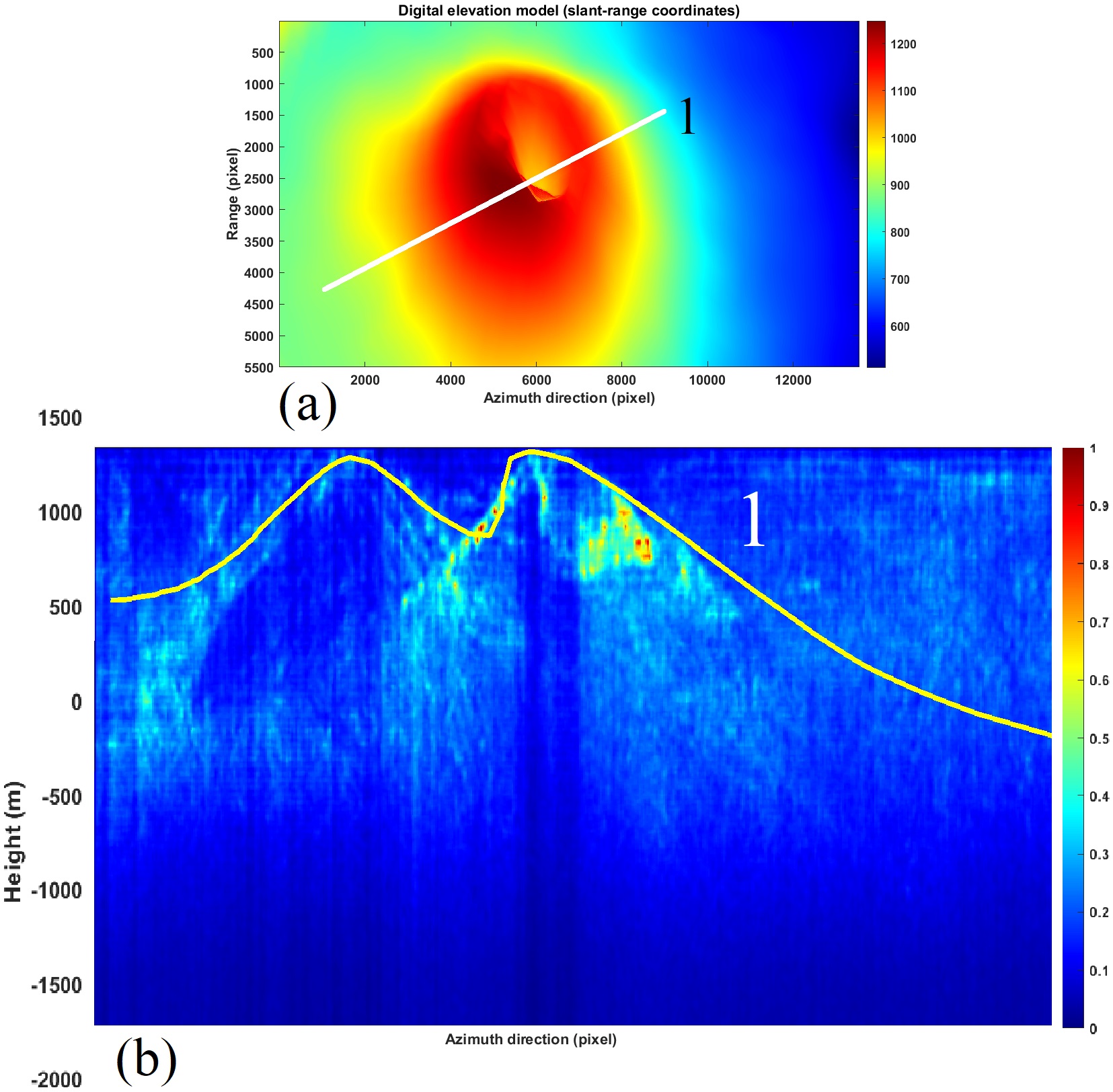}
\caption{Range-line echographic tomography result (magnitude), and validation through DEM. (a): DEM result in the slant-range coordinates. (b): Tomographic result (in magnitude), and topographic height (yellow line grabbed from the white line 1 of Figure (a)).}
\label{DEM_3}
\end{figure}

In this section, the topographic validation of the estimated tomograms is performed. The topographic elevation lines are provided by the digital elevation model (DEM) generated by the Shuttle Radar Topography Mission (SRTM). The topographic lines are superimposed on the estimated tomographic sections. In this specific case the topographic line 1 visible in Figure \ref{DEM_3} (a), has been extrapolated from the global SRTM topographic surface, which is coincident with the tomographic line 1 shown in Figure \ref{Obliqua_1} (a). The topographic line has been merged on the tomogram and the total information is visible in Figure \ref{DEM_3} (b). The visual analysis shows the surface tomogram overlapping almost perfectly the topographic line, which is represented by the yellow line 1.
	
\begin{figure}[htp]
 \centering
\includegraphics[width=13.0cm,height=14cm]{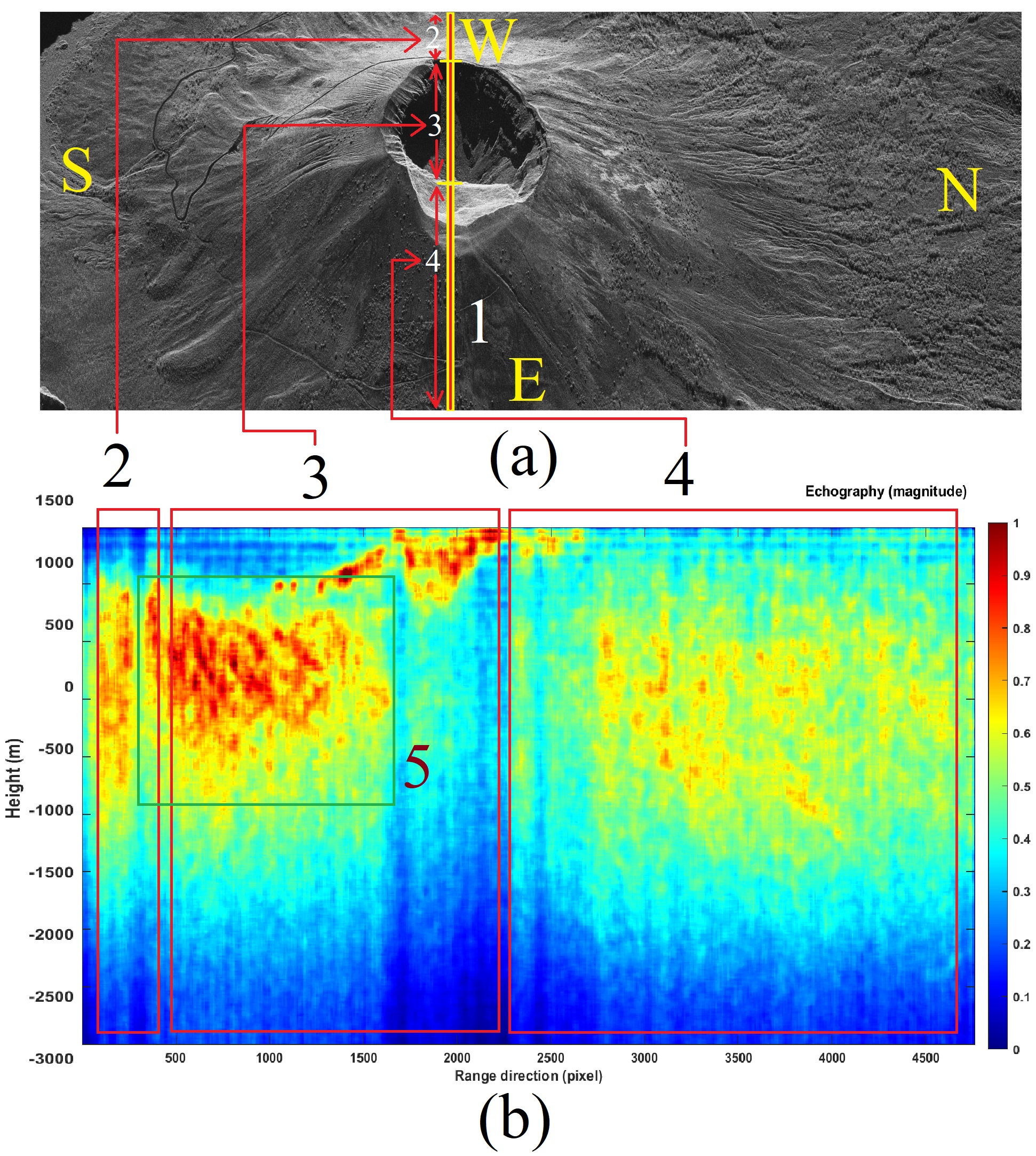}
\caption{Tomographic results and analysis. (a): SLC SAR data of the Vesuvius. (b): Tomographic results and analysis.}
\label{DDD_1}
\end{figure}

\begin{figure}[htp]
\centering
\includegraphics[width=12cm,height=9.0cm]{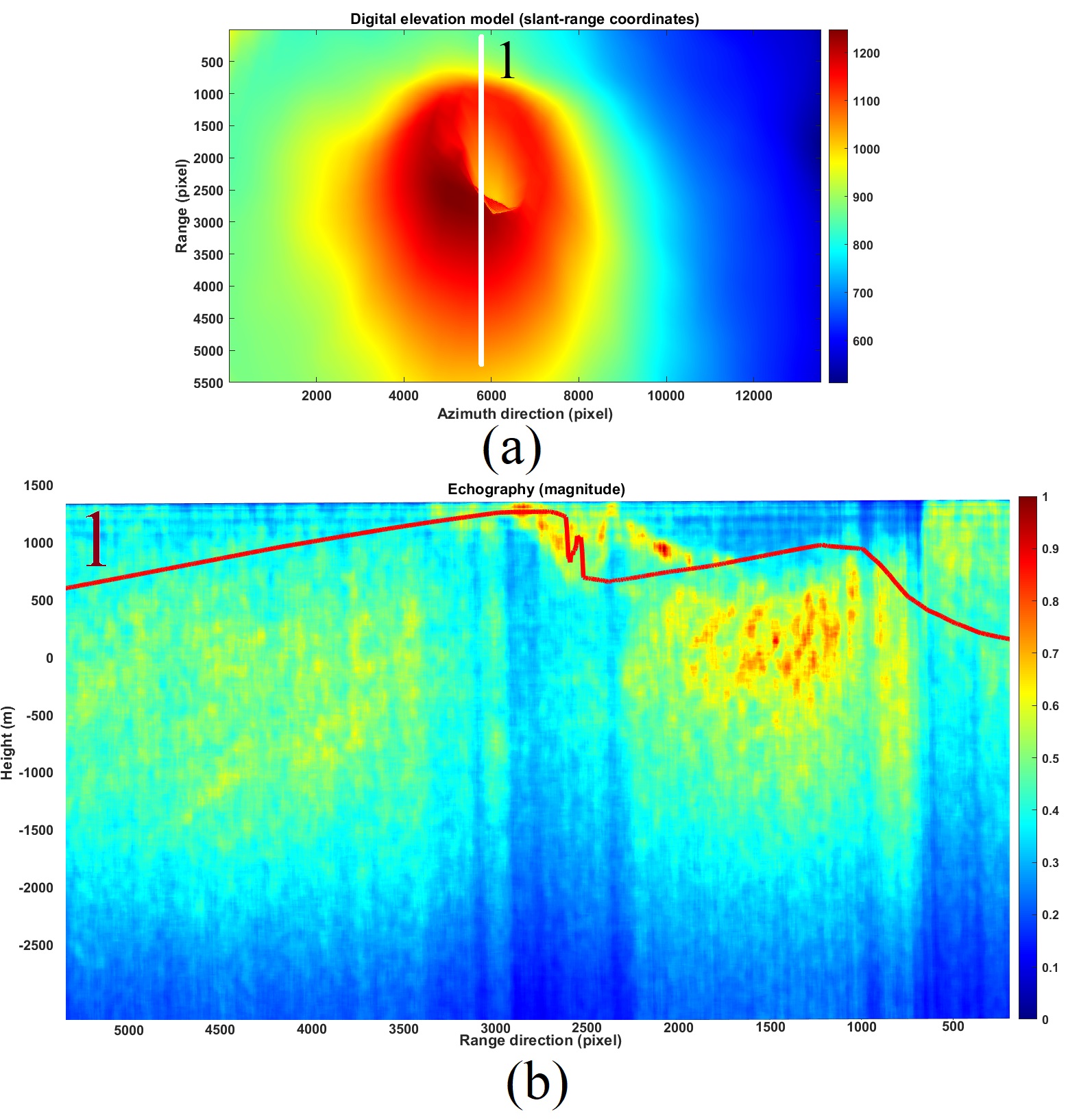}
\caption{Range-line echographic tomography result (magnitude), and validation through DEM. (a): DEM result in the slant-range coordinates. (b): Tomographic map (in magnitude), and topographic height (yellow line). The result is calculated over the white line drown on the DEM reported in Figure (a).}
\label{DEM_1_Range}
\end{figure}

The next validation case is to compare the position of the topographic line 1 visible in Figure \ref{DEM_1_Range} (a) (always extrapolated from the SRTM-DEM), with the top of the tomography coincident to line 1 shown in Figure \ref{Range_599_1} (a) and the consequent merged information is visible in Figure \ref{DEM_3} (b). The visual analysis shows the almost perfect overlap of the superficial component of the tomogram with the topographic line, which is represented by the yellow line 1. The last performance analysis case study is shown in Figure \ref{DDD_1} (a), (b). For this case, the tomographic line is perfectly oriented along the range direction. The objective was to study the tomographic reconstruction robustness under heavy foreshortening and layover effects. The tomographic picture is reported in Figure \ref{DDD_1} (b). In this case, the eastern side of the crater (the one characterized by layover) is better seen than the western side (the one characterized by foreshortening). However, possible lava conduits communicating with both the inner and outer parts of the main volcano cone remain visible on the eastern side of the crater. The tomogram has been divided into three portions: results numbers 2, 3, and 4, which are compartmentalized by three red squares visible in Figure \ref{DDD_1} (b). Each tomographic surface is pointed to its tomographic line reference, which is visible in Figure \ref{DDD_1} (a). The portion of the tomogram characterized by layover is the number 2, while result number 3 is the one representing the interior of the main crater. Finally, result number 4 refers to the descending side of the volcanic cone. Results confirm the presence of vibratory material, similar to rock, probably composed of cooled compact lava that produces a substantial blockage mass located deeply inside the main volcanic conduit. This blockage is visible within the box 5 of Figure \ref{DDD_1} (b).

\subsection{Vibration Validation}
Vibrational validation is carried out by comparing data estimated by the satellite with those observed by the "in-situ" seismological stations installed at Vesuvius. In-situ data is provided in open source from the Italian National Institute of Geophysics and Volcanology (INGV), and downloadable from the following website: https://eida.ingv.it/it/. In this context, the vibrational data estimated by the satellite and those provided by the seismic in-situ stations listed in Table \ref{Table_2}, are synchronized in space and time.
	
\begin{table}[tb!]
\caption{Characteristics of the "in-situ" seismographs.}
\begin{center}
\begin{tabular}{ p{4cm} p{2cm} p{2cm} p{4cm}}
\hline 
"In-situ" station & Network & Channel & Location (Lat, Lon)\\
\hline
Vesuvius - East Crater & IV-VCRE & HH & 40.818999$^\circ$, 14.431419$^\circ$\\
Vesuvius - North Bunker & IV-VBKN & HH & 40.829959$^\circ$, 14.429881$^\circ$\\
\hline
\end{tabular}
\label{Table_2}
\end{center}
\end{table}
	
\begin{figure}[htp]
\centering
\includegraphics[width=11.0cm,height=8.0cm]{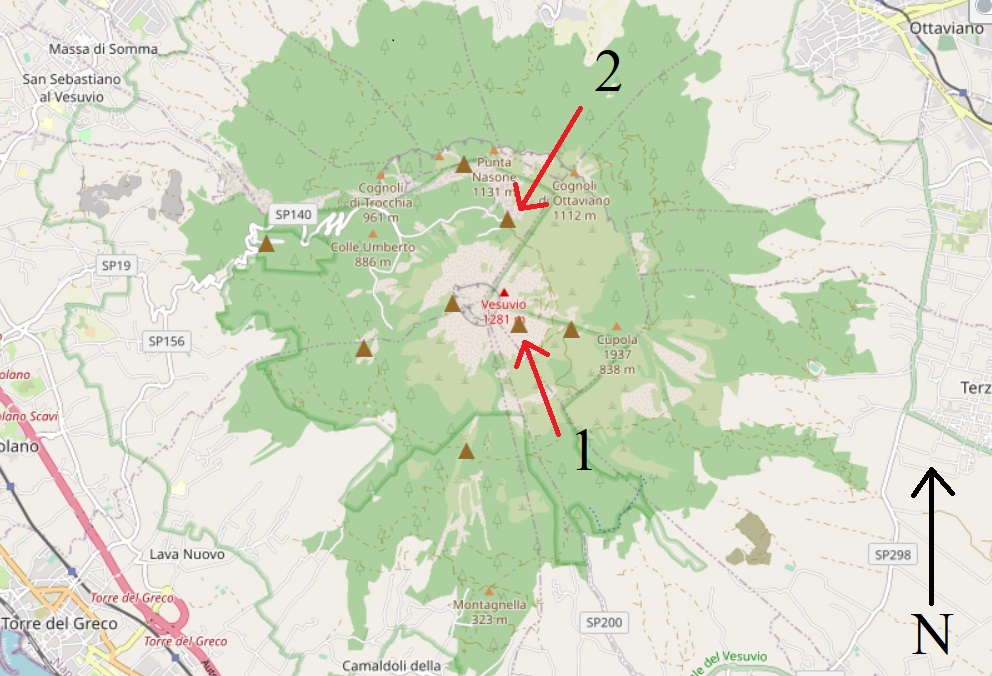}
\caption{"In-situ" seismograph stations.}
\label{Stazioni_1}
\end{figure}

\begin{figure}[htp]
\centering
\includegraphics[width=14.0cm,height=5.0cm]{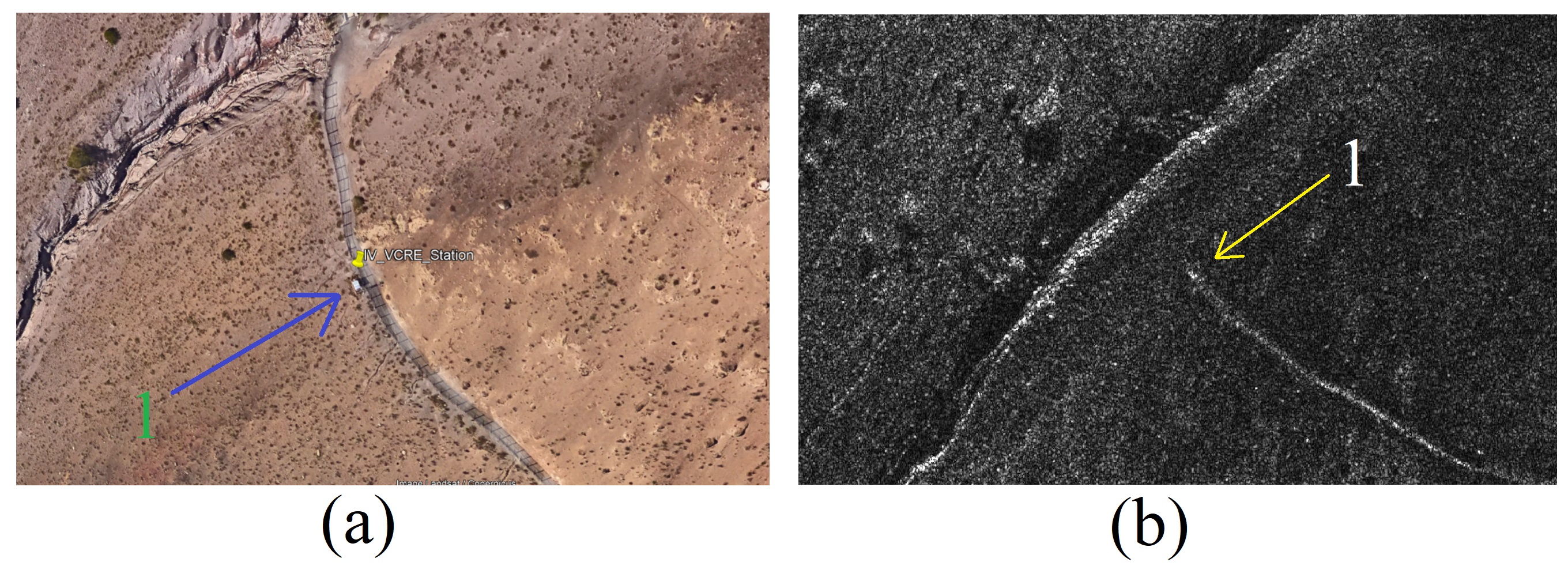}
\caption{IV-VCRE seismograph station, located on the East side of the crater.}
\label{In_Situ_1}
\end{figure}

\begin{figure}[htp]
\centering
\includegraphics[width=14.0cm,height=5.0cm]{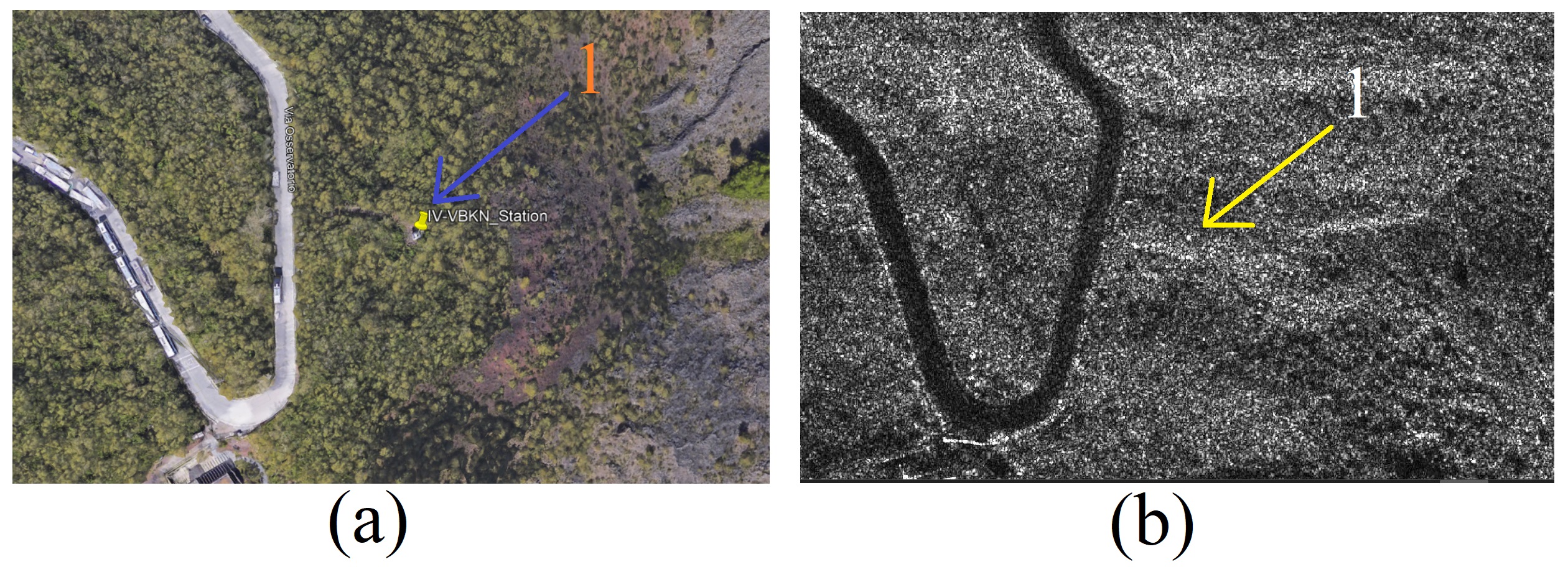}
\caption{IV-VBKN seismograph station, located on the North side of the crater.}
\label{In_Situ_2}
\end{figure}

\begin{figure}[htp]
\centering
\includegraphics[width=15.0cm,height=5.0cm]{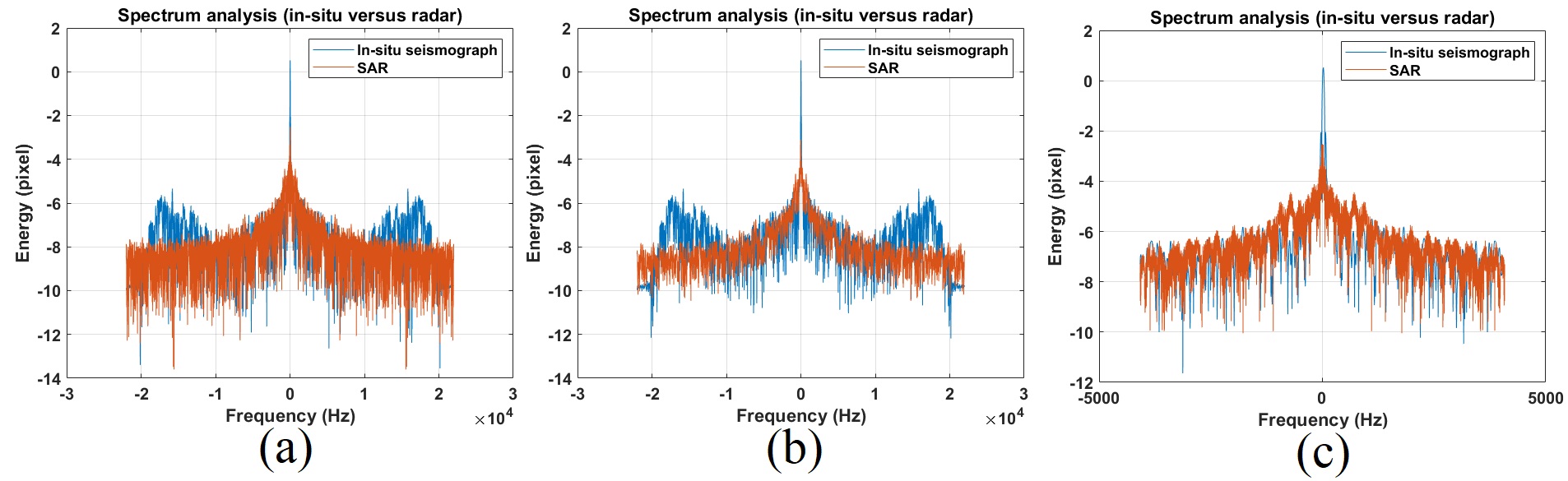}
\caption{IV-VCRE seismograph station versus SAR frequency-domain vibrational streaming. (a): Native (b): low-pass filtered. (c): Narrow-band particular.}
\label{VCRE_Spectrum_1}
\end{figure}

\begin{figure}[htp]
\centering
\includegraphics[width=15.0cm,height=5.0cm]{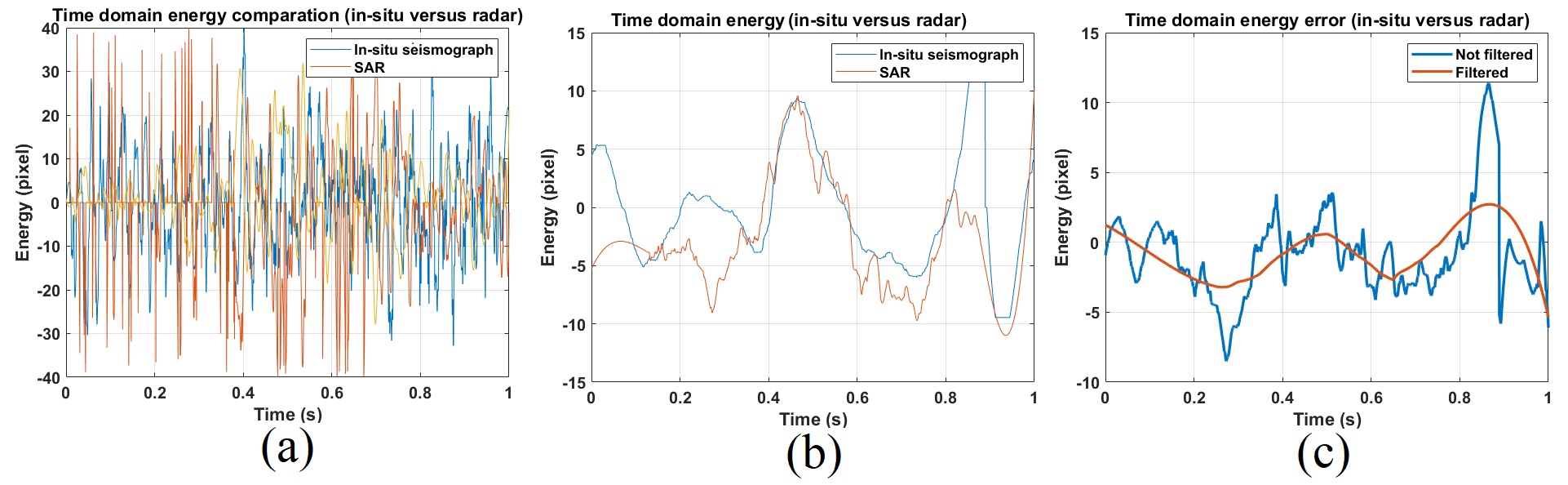}
\caption{IV-VCRE seismograph station versus SAR synchronized time-domain vibrational streaming. (a): Native (b): low-pass filtered. (c): errors.}
\label{VCRE_Time_Domain_1}
\end{figure}

\begin{figure}[htp]
\centering
\includegraphics[width=15.0cm,height=5.0cm]{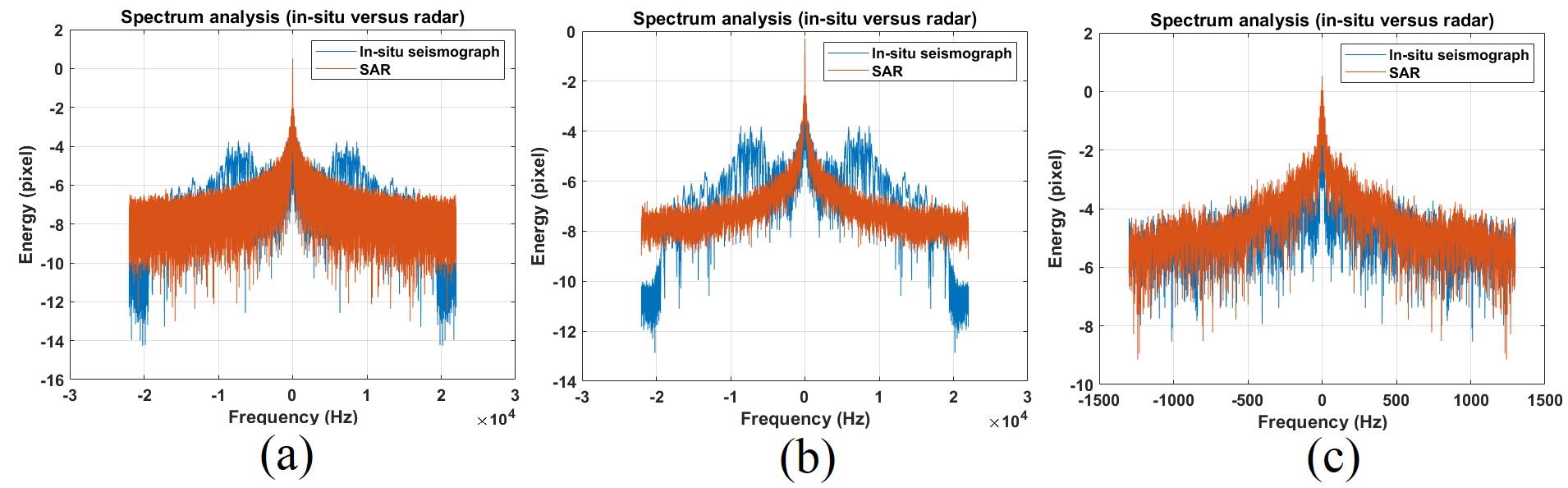}
\caption{IV-VCRE seismograph station versus SAR synchronized signals spectrum. (a): unfiltered. (b): low-pass filtered. (c): 1kHz filtered. particular}
\label{VBKN_Spectrum_1}
\end{figure}

\begin{figure}[htp]
\centering
\includegraphics[width=13.0cm,height=5.0cm]{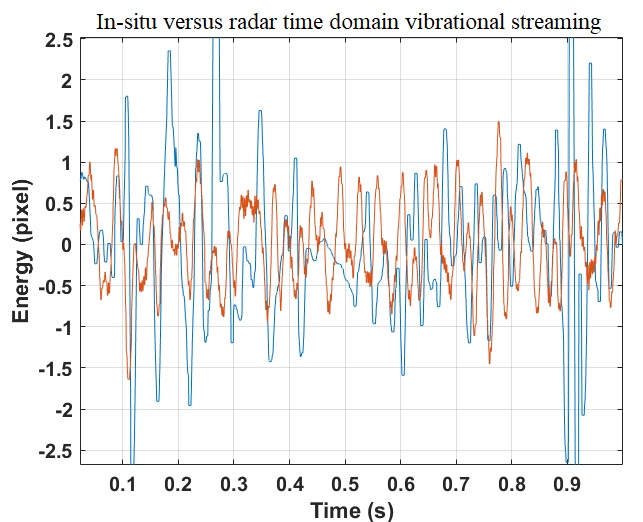}
\caption{IV-VCRE seismograph station versus SAR synchronized time-domain vibrational streaming.}
\label{VBKN_Time_Domain_1}
\end{figure}

Figure \ref{Stazioni_1} shows the map where the Vesuvius seismic stations are located. IV-VCRE and VBKN in-situ station data are considered in this stage and are the only compatible with the line-of-sight radar visibility. All other measurement stations were either not visible or located outside the SAR imaging scene. Figures \ref{In_Situ_1} (a), and (b), and Figure \ref{In_Situ_2} (a), and (b), represents the optical and radar observations, respectively, of the seismographic stations, The first facility is located at the east side of the crater and the second one is installed toward the northern side. The two stations are successfully represented on both the optical and radar images (indicated both by arrows number 1). In Figure \ref{VCRE_Spectrum_1} (a), (b), and (c), the vibrational spectrum streaming recorded by the IV-VCRE seismographic station, superimposed to the respective spectrogram recorded by radar, is depicted. Figure \ref{VCRE_Spectrum_1} (a) represents the unfiltered full-bandwidth spectrum, while Figure \ref{VCRE_Spectrum_1} (b) is the filtered spectrum, finally Figure \ref{VCRE_Spectrum_1} (c) represents a narrowed detail spectrum at approximately 1 kHz of bandwidth. The blue color function is that of the "in-situ" station, while the brown color function is the radar-estimated vibrational spectrum. The native time trend (1 second of synchronized streaming) is shown in Figure \ref{VCRE_Time_Domain_1} (a). The filtered time-domain plot (approximately composed of 1-second wide synchronized data streaming), is reported in Figure \ref{VCRE_Time_Domain_1} (b), finally the IV-VCRE in-situ station versus SAR vibrational streaming measurement errors is proposed in Figure \ref{VCRE_Time_Domain_1} (c). The blue function represents the unfiltered error while the one in brown is the filtered error. It can be seen that this error while oscillating around zero, remains confined to very low values. In Figure \ref{VBKN_Spectrum_1} (a), (b), and (c), the spectrum of the vibrational streaming recorded by the IV-VBKN seismographic station is represented. Figure \ref{VBKN_Spectrum_1} (a) represents the spectrum in its entirety, Figure \ref{VBKN_Spectrum_1} (b) is the filtered spectrum while Figure \ref{VBKN_Spectrum_1} (c) represents a narrowed detail datasets at approximately 1 kHz of bandwidth. Also in this case the blu and the brown functions are the in-situ and the radar estimated vibrational spectrum respectively. The unfiltered time-domain sampled vibrations (approximately 1-second synchronized streaming) are shown in Figure \ref{VBKN_Time_Domain_1}. Although it is very difficult to perfectly synchronize two different measurement instruments, the first one located on the ground, while the second one is located in space, such time synchronization turns out to be very acceptable. Both instruments are measuring the seismic background of the area. This result brings to the attention of the SAR's effective measurement accuracy.
	
\subsection{Tomographic Validation}
\begin{figure}[htp]
\centering
\includegraphics[width=12.0cm,height=10.0cm]{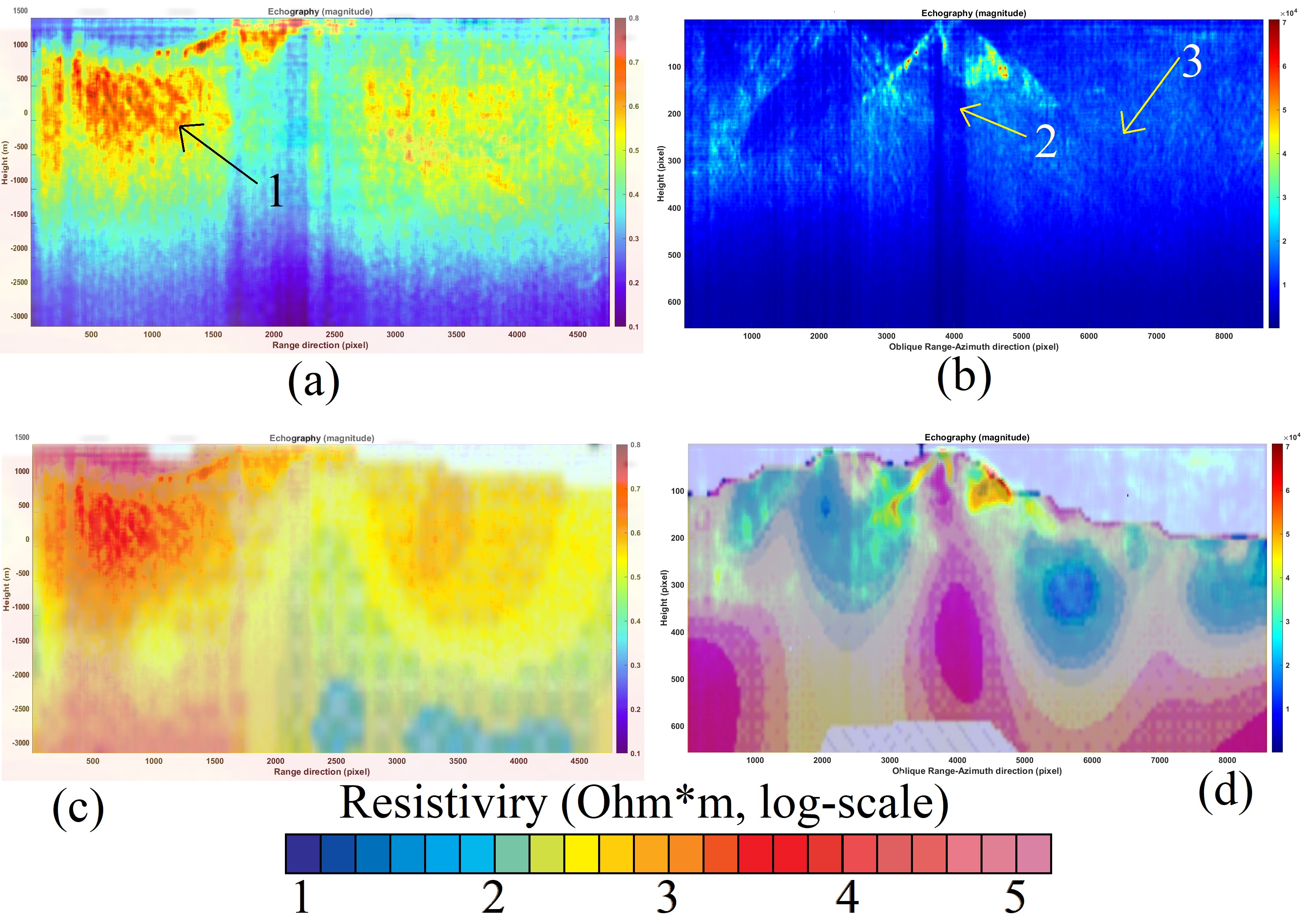}
\caption{Overlay of SAR Doppler tomography with magnetotelluric tomography.}
\label{Magnetotelluric_2}
\end{figure}

\begin{figure}[htp]
\centering
\includegraphics[width=14.0cm,height=16.9cm]{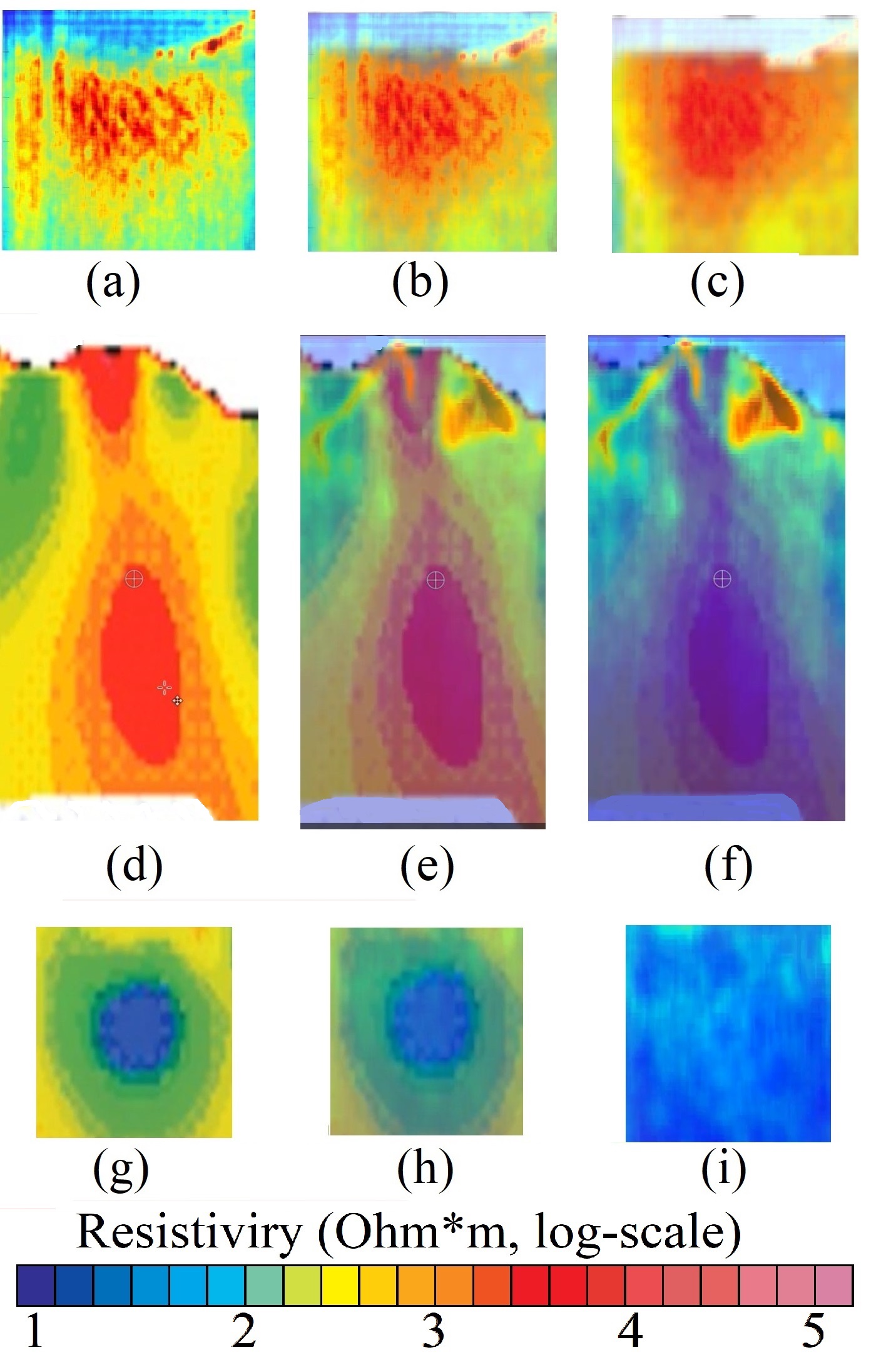}
\caption{Overlay of SAR Doppler tomography with magnetotelluric tomography (particulars).}
\label{Magnetotelluric_1}
\end{figure}

\begin{figure}[htp]
\centering
\includegraphics[width=14.0cm,height=14.0cm]{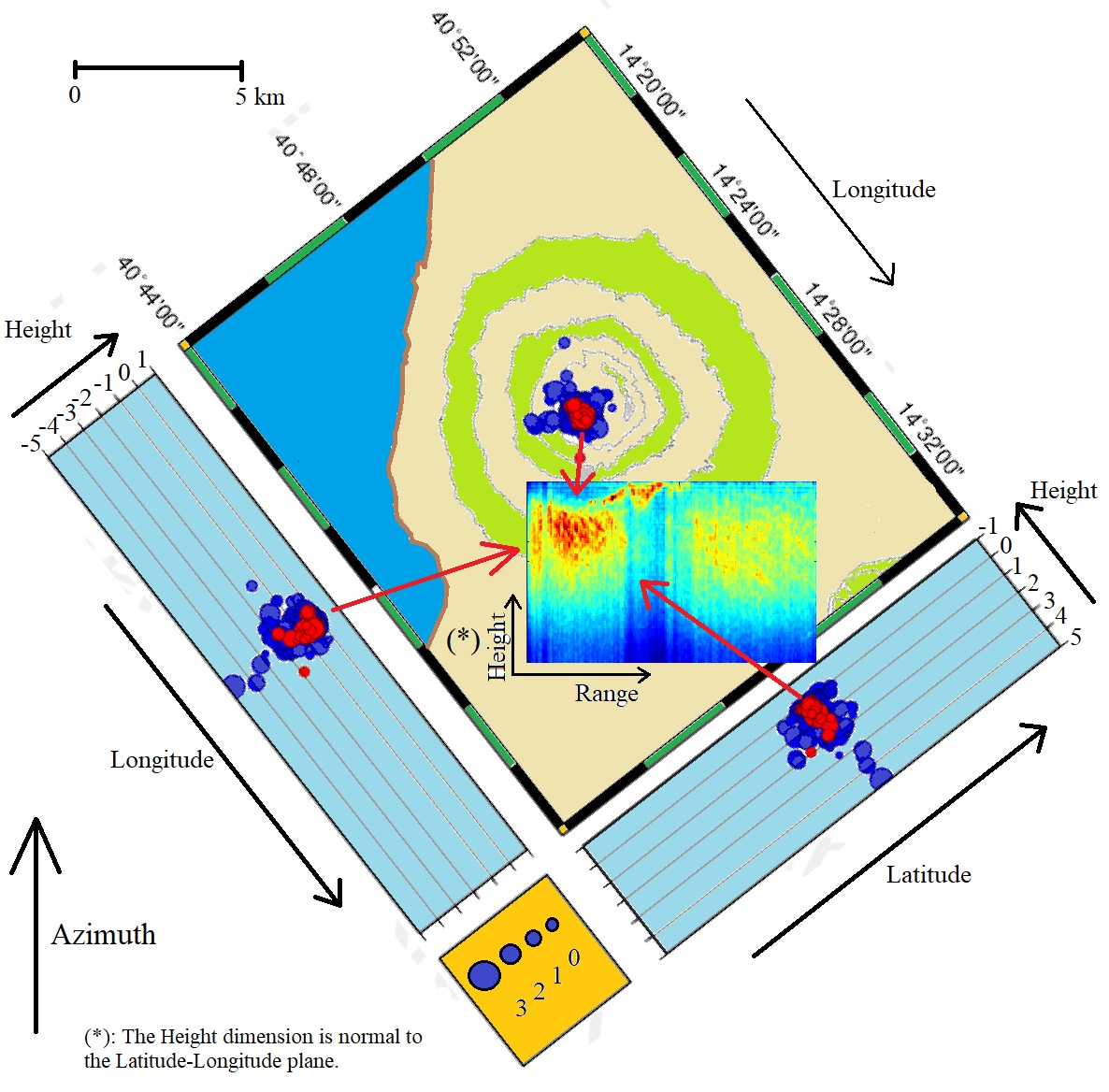}
\caption{Cartographic representation of the seismic events that occurred during the year 2022 at Vesuvius. The pegs represent in the position the one actually occurred in 3D space (latitude-longitude and depth), the radius the magnitude. The red roundels represent those occurred the month where the SAR acquisition was taken. The magnitude legend has four scales of earthquake in the Richter scale, from the smallest (0 to 1) to the largest (greater than 3).}
\label{Fig_All_2}
\end{figure}

In this section, the comparison of data estimated through magnetotelluric tomography with those estimated through SAR tomography is evaluated. Successively the visual comparison relates the magnitude of the earthquakes that occurred during the month, within which the SAR observation is made. The estimated tomographic planes found the perfect overlap of the estimated results with the radar. In Figure \ref{Magnetotelluric_2} (a), (b), (c), and (d) we report the radar-estimated tomographic maps superimposed on the magnetotelluric tomography results. More precisely, Figure \ref{Magnetotelluric_2} (a) is the in-range SAR tomography, Figure \ref{Magnetotelluric_2} (c) is the superposition of Figure \ref{Magnetotelluric_2} (a) radar tomography with magnetotelluric tomography. On the other hand, Figure \ref{Magnetotelluric_2} (b) shows the range-azimuth oblique SAR tomography, while Figure \ref{Magnetotelluric_2} (c) is the superposition of the oblique radar tomography in Figure \ref{Magnetotelluric_2} (b) with the magnetotelluric tomography. In Figure \ref{Magnetotelluric_2} (a, i) some details are depicted to better show the existing tomographic overlay. The images in Figure \ref{Magnetotelluric_2} (a), (b), and (c) refer to the topographic detail of Figure \ref{Magnetotelluric_1} (a), where a reference area is indicated by arrow number 1. Figure \ref{Magnetotelluric_2} (a) is the radar data alone, Figure \ref{Magnetotelluric_2} (b) is the 50\% overlay of the radar data on the magnetotelluric tomography, and finally Figure \ref{Magnetotelluric_2} (c) is the magnetotelluric tomography map showed alone. The images of Figure \ref{Magnetotelluric_2} (d), (e), and (f) are referred to the tomographic detail of Figure \ref{Magnetotelluric_1} (b), indicated by arrow 2. Figure \ref{Magnetotelluric_2} (d) is the radar data alone, Figure \ref{Magnetotelluric_2} (d) is the 50\% overlay of the radar data with the magnetotelluric tomography, and finally Figure \ref{Magnetotelluric_2} (e) is the magnetotelluric tomography, again showed alone.
Finally, the images of Figure \ref{Magnetotelluric_2} (g), (h), and (i) are referred to the tomographic detail of Figure \ref{Magnetotelluric_1} (b), indicated by arrow 3. Figure \ref{Magnetotelluric_2} (g) is the radar data alone, Figure \ref{Magnetotelluric_2} (h) is the 50\% overlay of the radar data with the magnetotelluric tomography, and finally Figure \ref{Magnetotelluric_2} (i) is the magnetotelluric tomography without any overlay.
The comparative assessment formed by the visual validation between the results estimated through the seismic sensors network installed in the proximity of Vesuvius, and measured satellite results are secondly performed. The data have been extrapolated from the public and institutional websites of the INGV. In this context, the seismic values map is shown in Figure \ref{Fig_All_2} which is correlated to the SAR tomographic results of Figure \ref{DDD_1} (b). In the Figure, blue and colored circles represent some seismic events. The center of each point is placed in the 3D space on which the source of the seismic event was measured, while the radius of each circle measures its magnitude. The Figure also contains the magnitude legend and four scales can be distinguished, the number 0 value represents all seismic events between 0 and 1 Richter scale magnitude. The number 1 represents all events of magnitude between 1 and 2, the circle 2 between 2 and 3, and finally, the largest circle is the number three which represents all seismic events greater than 3 degrees of magnitude on the Richter scale. The blue circles belong to represent seismic events that occurred throughout January 2022, while the red color represents only those seismic events that occurred in February 2022, which is the month in which the SAR data was acquired. From a visual comparison, we keep a good correspondence of the data if compared to the one estimated through SAR tomography. It is clear that the temporal correspondence of the SAR acquisition which consists of a few seconds is not comparable with the large temporal duration of the measurements made through the "in-situ" sensors, but we find a good correlation between the red seismic events that coincide very much in space with the magmatic consistency estimated through the SAR seismic tomography. The depth also coincides greatly.

\section{Discussion}\label{Discussion}
This is the first time that a SAR has been employed to estimate the consistency of soil depth, down to 3 km, considering also that the observations are performed from space. This preliminary work may pave the way for a new type of radar exploitation, where the carrier physical phenomenon is formed by coherent electromagnetic X-band transmissions and azimuth focusing made using ad-hoc designed matched filters tuned at the zero-Doppler. This procedure can grab phononic (intended to be the vibration of matter) physical parameters. In this context, photons are used as a carrier medium that contains phononic information as well. It seems that this system, although still to be improved and refined, works. This technique can be extended to maybe detect crude-oil, or natural gas underground pockets, to quickly search for veins in metals and rare earth, or to assess the consistency of the matter from which all the world's great infrastructures are made. In the present work, we first thought about both nature and humanity preservation, and then found a method (which at present remains unresolved), namely that of looking for inside volcanoes with high-resolution imaging from space. It seemed natural to us to focus our research on one of the world's most dangerous volcanoes. It is located in Italy in the middle of the highly-populated city of Naples, the Vesuvius. In addition, this technique allows the construction of an accurate and truthful model of the Earth's subsurface. This possibility appears very important and could serve in helping to strengthen predictive models useful in both the volcanological and seismological fields.
According to \cite{biondi2019atmospheric}, there is an electromagnetic interaction with the atmosphere. The technique employs the single SAR image, acquired in a time of about 14 seconds, surely there is an electromagnetic phase delay due to the atmosphere. But this delay, in practice, does not affect the correct estimation of the tomography, as it is constant in time (this effect is assumed to be time-invariant), so we assume that atmospheric delay remains constant within the SAR acquisition. However, the proposed technique is very robust to compensate for atmospheric interaction as we scan in the Doppler Domain, within the single SAR image. To this end, MCA of atmospheric issues is fully described and solved in \cite{biondi2020monitoring}.
Concluding, the proposed technique, can be considered a potential "gap-filler", thus allowing us to look inside rigid bodies, like volcanoes, even over high spatial resolution. Authors employed specific designed software for processing tomographic slices. At the present, there is no commercial software capable of extracting the phonon information embedded in the SAR data. The authors are available in collaborating with other research groups for reproducing the proposed tomographic method specifically for all those sites that need to be studied in depth.
	
\section{Acknowledgements}\label{Acknowledgements}
We would like to thank Prof. Daniele Perissin for making the SARPROZ software available, through which many calculations were carried out more easily and quickly. We also thank the Italian Space Agency for providing the SAR data. The in-situ seismic numerical data were kindly provided by the Italian Institute of Geophysics and Volcanology (INGV), which can be downloaded from the website: https://eida.ingv.it/it/. Finally, We thank Prof. Corrado Malanga, Professor of Industrial Chemistry at the Faculty of Chemistry, University of Pisa (Italy), for guiding me in formulating the idea of extrapolating phononic information through photonic processing of SAR data. (https://corradomalangaexperience.com/).
	
\section{Conclusions}\label{Conclusions}
This work describes an imaging method based on the analysis of micro-motions present on volcanoes, and generated by the underground earth tremor.
We showed series of tomographic maps representing the internal echography of volcanoes. Processing the coherent vibrational information present on the single SAR image, in the single-look-complex configuration, we exploited the sound information penetrating tomographic imaging over a depth of about 3 km from the Earth's surface. The experimental results were calculated processing a SLC image from the COSMO-SkyMed Second Generation satellite constellation of the Italian volcano Vesuvius.
Tomographic maps revealed the presence of the magma chamber, the main and the secondary  volcanic conduit. In addition, both the main and secondary vents are visible.
	
\section{Author contributions statement}
The authors contributed to all parts of this work

\bibliographystyle{unsrt}  
\bibliography{references}

\end{document}